# Moral Decision-Making in Medical Hybrid Intelligent Systems

## A Team Design Patterns Approach to the Bias Mitigation and Data Sharing Design Problems

Jip van Stijn

MSc Thesis Artificial Intelligence

Vrije Universiteit Amsterdam – Annette ten Teije

TNO – Mark Neerincx

# Abstract

Increasing automation in the healthcare sector calls for a Hybrid Intelligence (HI) approach to closely study and design the collaboration of humans and autonomous machines. Ensuring that medical HI systems' decision-making is ethical is key. The use of Team Design Patterns (TDPs) can advance this goal by describing successful and reusable configurations of design problems in which decisions have a moral component, as well as through facilitating communication in multidisciplinary teams designing HI systems. For this research, TDPs were developed to describe a set of solutions for two design problems in a medical HI system: (1) mitigating harmful biases in machine learning algorithms and (2) sharing health and behavioral patient data with healthcare professionals and system developers. The Socio-Cognitive Engineering methodology was employed, integrating operational demands, human factors knowledge, and a technological analysis into a set of TDPs. A survey was created to assess the usability of the patterns on their understandability, effectiveness, and generalizability. The results showed that TDPs are a useful method to unambiguously describe solutions for diverse HI design problems with a moral component on varying abstraction levels, that are usable by a heterogeneous group of multidisciplinary researchers. Additionally, results indicated that the SCE approach and the developed questionnaire are suitable methods for creating and assessing TDPs. The study concludes with a set of proposed improvements to TDPs, including their integration with Interaction Design Patterns, the inclusion of several additional concepts, and a number of methodological improvements. Finally, the thesis recommends directions for future research.

# Table of Contents

# 1. Introduction

## 1.0 Introduction

Over the past decades, the healthcare domain has witnessed a steep increase in automation. eHealth applications allow for higher quality and more cost-effective care (Elbert et al., 2014), robot-assisted surgery has proven to be effective and safe in several medical domains (Ghezzi & Corleta, 2016), and machine learning algorithms are capable of classifying radiology images with malicious cancers better than many a radiologist (Lakhani et al., 2018). As cutting-edge technological advancements continue to be made, there is ample reason to believe that the near future will continue to bring increasingly autonomous systems into the medical sector. While this process of automation has unprecedented potential, concerns have been voiced both in the academic realm and society. Dystopian scenarios sketch autonomous systems slowly replacing healthcare professionals, providing cold and impersonal care to patients. Others point at undesirable effects of current autonomous systems, such as the death of pedestrian Elaine Herzberg due to a collision with a self-driving car (Heaven, 2018), deaths resulting from the use of surgical robots (Alemzadeh et al., 2016), or algorithms maltreating minorities in their job chances (Dastin, 2018), access to healthcare (Obermeyer et al., 2019), and even exam results (Adam, 2020). It is undoubtable that medical autonomous systems are bound to make mistakes that influence life and death while they, unlike human health care professionals, cannot be held accountable for their actions (Pepito et al., 2019).

These are legitimate worries that should be taken into serious consideration. Due to the high beneficial and capitalist potential of autonomous systems, it seems implausible and unreasonable to stop the movement of increasing automation in healthcare altogether. What is possible, however, is to study the positive and negative effects of these systems, and design them carefully as to make sure that their actions align with our goals and values.

## 1.1 State of the art

While it is certain that the job description of many healthcare professionals will change with the technological advancements, the fear that autonomous systems will completely take over the healthcare sector has diminished substantially in recent years. There is a general academic consensus that, albeit we may not be able to fully apprehend what artificial intelligence will be capable of in the far future, humans possess a distinct set of qualities that machine agents will not be able to parallel in the near future (Dellermann et al., 2019). Among such qualities are thinking outside the box, reacting to unexpected situations, and navigating the social world. In fact, a growing body of literature advocates that the abilities of human and artificial intelligence should not be viewed as competitive, but rather as complementary. This philosophy has coined the term 'Hybrid Intelligence' (HI), aiming at utilizing the complementary strengths of human and artificial intelligence, so that they can perform better than either of the two separately (Kamar, 2016; Peeters et al, 2020; Akata et al., 2020).

Following this paradigm implies studying not just the cooperation of humans with an autonomous system, but rather the union of humans and machine agents in a single system. This conception allows for research that would not be possible from either the autonomous or human factors perspectives (e.g. studying the interactions between the weaknesses and strengths of both types of agents). The field of hybrid intelligence comes with its own set of unique challenges which are inherently interdisciplinary in nature. One of these challenges is the study of how complex human-computer teams take moral decisions, and, following from this, how we can design such systems in a way that they behave ethically. As the previous section pointed out, there is a pressing demand for research in this area, considering the fact that moral decisions in medical human-computer teams are already affecting many people and will soon affect many more, having crucial consequences for their health.

The moral component of autonomous machines has been the subject of recent studies, especially in the field of machine ethics. This discipline attempts to contribute to the creation of Artificial Moral Agents (AMAs), that follow certain ethical rules, either implicitly or explicitly (Anderson & Anderson, 2007). However, most of this research focuses on fully autonomous systems, excluding the human component of how these machines are used in practice. From a hybrid intelligence approach, it is key to not only include machine requirements but also human cognitive capabilities in the study of AMAs. As this is a highly interdisciplinary endeavor, it is first and foremost crucial to establish a common language to talk about moral situations in Hybrid Intelligent systems. Van der Waa et al. (2020) have made a first attempt using so-called Team Design Patterns (TDPs), describing the task allocation of human-computer teams in morally sensitive situations. However, this language has minimal sufficiency for expressing human cognitive components and requirements. This is a vital element for a truly hybrid intelligent approach toward moral decision making in autonomous systems. Additionally, the method has so far solely been used as a taxonomy and has not been utilized in the design of a system, meaning that there have been no evaluations of its effectiveness.

## 1.2 Research aims

The aim of the current thesis is twofold. Firstly, it attempts to advance the conceptualization of moral decision-making in hybrid intelligent systems in (and perhaps even outside) the medical domain. It will do so by creating Team Design Patterns for two use cases: bias mitigation and data sharing in a hybrid intelligent digital assistant for diabetes type II care. These patterns are a contribution to the academic literature on their own, as they are reusable entities for solving similar design challenges. Additionally, the creation of these patterns is meant to contribute to filling the gaps in the literature mentioned above: (1) the expression of both human cognitive components and AI requirements in the pattern language and (2) the development of TDPs not only as a library of successful and reusable design solutions, but also in their application as a method for the design process Hybrid Intelligence by a multidisciplinary team.

Secondly, this thesis aims to contribute to a scientific standard in the methodology of conceptualizing moral decision-making in hybrid intelligent systems, as there is currently no such standard yet. This research will carefully document the used methods both for arranging the TDPs as well as their usability evaluation. As there are currently no documentations of methods for creating and evaluating TDPs, it is outside the scope of this thesis to fully establish such a methodology. Rather, it may serve as a first benchmark in the development of a sound methodological framework for the conceptualization of moral decision-making in human-computer teams.

## 1.3 Research questions

Following from the research aims, the current thesis attempts to answer several research questions (RCs):

> *RQ 1: How can Team Design Patterns describe moral decision-making for bias mitigation and data sharing in a medical hybrid intelligent system, so that they are usable by researchers from the various disciplines involved in the design of such systems?*

As will become clear in Chapter 4, usability of the Team Design Patterns is defined as having three main components: understandability, generalizability, and effectiveness. Hence, the current thesis will answer three sub-questions of the first research question.

> *RQ 1a: How understandable are the created Team Design Patterns for their potential users?*
> *RQ 1b: How generalizable are the created Team Design Patterns?*
> *RQ 1c: To what extent does the use of the created Team Design Patterns lead to more appropriate moral decision-making?*

Additionally, a second research question is defined for the research aim of developing a standard methodology for creating Team Design Patterns:

> *RQ 2: Which methodological tools are suitable for the creation of such Team Design Patterns?*

The next section gives an overview of how the current thesis is organized to provide an answer to these research questions.

## 1.4 Contents

The second chapter of this thesis describes the methodology and research design used in this research. It first addresses the two methodological theories the design of this research is based on: Socio-Cognitive Engineering (SCE) and Value-Sensitive Design. It then describes the research design, including the use-case of a Hybrid Intelligent system to counter Diabetes Type II through prevention, diagnosis, treatment, and management.

The third chapter presents the foundation component from the SCE approach. It starts with an outline of the operational demands, analyzing what kind of support is needed in the moral domain of the system for diabetes care. This leads to the identification of two main design challenges: the mitigation of harmful biases in learning models and the sharing of patient data for medical and research aims. Subsequently, it analyzes the relevant human factors, focusing on psychological theories of moral decision-making. Finally, it gives an overview of the technological principles relevant to the envisioned system, including the distinction between machine learning and knowledge reasoning, and an overview of the current academic literature regarding techniques and solutions to the design challenges.

Chapter 4 consists of the specification phase of the research, explaining how the knowledge from the foundation component was integrated into a set of patterns. This chapter addresses the objectives, functions, and claims of the design patterns, after which the proposed Team Design Patterns are presented.

Chapter 5 describes the setup of the evaluation of the design patterns proposed in chapter 4. It suggests the development and use of a survey to obtain a combination of qualitative and quantitative data from researchers with varying backgrounds. It then presents the qualitative and quantitative results of this questionnaire.

Finally, Chapter 6 discusses the findings of this research, and provides suggestions for improvements of the team design patterns as presented in Chapter 4. Additionally, this chapter reflects on the methodology of this research. Finally, this chapter concludes with a brief summary of the research and a set of suggestions for future research in the area of moral decision-making in medical Hybrid Intelligent systems.

# 2. Methodology & Research Design

## 2.0 Introduction

The previous chapter demonstrated the need for studying moral decision-making in hybrid intelligent systems and presented a set of research questions to advance this goal. The current chapter sets out the methodological framework for this thesis, and the research design following from adopting this methodology. The first section (2.1) gives a detailed explanation of Team Design Patterns, including an example scenario. It then discusses the Socio-Cognitive Engineering methodology and its integration with Value-Sensitive Design. The second section discusses the research design of the thesis, consisting of three phases: foundation, specification, and evaluation. Additionally, it presents the use case of a digital assistants for patients and healthcare professionals in the prevention, diagnosis, treatment and management of diabetes type II.

## 2.1 Methodology

As introduced in chapter 1, this thesis aims to conceptualize and describe moral decision-making in hybrid intelligent systems in the medical domain. A promising format for this is the use of Team Design Patterns (Van Diggelen & Johnson, 2019; Van Diggelen et al., 2019; Van der Waa et al., 2020): a combination of text and pictorial language to describe possible solutions to recurring design problems. These patterns represent reusable and generic HI design solutions in a coherent way aimed at facilitating the multidisciplinary HI design process. Subsection 2.1.1 presents the aims, use, and inner workings of Team Design Patterns as described in the academic literature so far. As there is no set of methods yet to arrive at Team Design Patterns, subsection 2.1.2 provides the motivation for adopting Socio-Cognitive Engineering methodology to create the desired design patterns, extended with elements of Value-Sensitive Design (Neerincx et al., 2019; Harbers & Neerincx, 2017).

### 2.1.1 Team Design Patterns

The notion of design patterns originated in the domain of architecture (Alexander et al., 1977). Alexander noticed that common design problems usually resulted in similar solutions and argued for a universal language to describe generic solutions for frequently occurring problems. For example, the design of an upstairs bathroom may be different for each specific house, but always requires sufficient foundation, piping, and mechanisms for safe use of electricity, which often results in the same type of design. Alexander (ibid.) stressed that these patterns should only capture the core of the solution on a general and abstract level to allow for the specific chances and limitations of each individual design. This way, *'you can use this solution a million times over, without ever doing it the same way twice'* (ibid.). Since then, design pattern languages have been developed in several other disciplines including workflow engineering, object-oriented programming, and interaction design (Van Diggelen & Johnson, 2019). Although the languages differ strongly across domains, they are often aided with graphical illustrations in order to optimally communicate the proposed solution of each pattern.

Recently, several scholars have attempted to create a design pattern language in the domain of human-agent teaming. However, an often-voiced worry in this domain is that it is difficult to codify the intuitive, straightforward process that is teamwork. Additionally, human-agent teamwork is relevant in many application domains with diverse sets of stakeholders, who all have differing jargon and varying knowledge of human-agent teaming. Van Diggelen and Johnson (2019) created so-called 'Team Design Patterns' (TDPs) with the aforementioned challenges in mind, resulting in a promising language to express human-agent teamwork. According to the authors, there are four requirements to these TDPs:

> 'Team pattern design solutions should be (1) *simple* enough to provide an intuitive way to facilitate discussions about human- machine teamwork solutions among a wide range of stakeholders including non-experts, (2) *general* enough to represent a broad range of teamwork capabilities, (3) *descriptive* enough to provide clarity and discernment between different solutions and situations, and (4) *structured* enough to have a pathway from the simple intuitive description to the more formal specification.'
>
> Van Diggelen & Johnson, 2019, p.118

Table 1 shows some of the main concepts of the pattern language, including their equivalents in the pictorial language. It uses images of actors (either a *human* or *machine agent*) carrying blocks with text to signifying the tasks each actor is responsible for. The boxes, or tasks, can either have a *direct* contribution to the team goal (indicated by opaque boxes) or an *indirect* contribution (aimed at making the team more efficient but not directly contributing to the team goal, indicated by partly transparent boxes). Patterns can consist of multiple *frames*, each showing a different allocation of tasks. A transition between these frames is depicted by solid arrows. Sometimes these transitions are initiated by a specific actor, which is indicated by a dotted arrow from that actor to the solid transition arrow. Van der Waa et al. (2020) attempted to use this pattern language specifically for describing task allocation in moral decision-making and noted that two other concepts were important. Firstly, they added a distinction between *moral* tasks, requiring some sort of moral capabilities (depicted with a red color) and *non-moral* tasks, not requiring such capabilities (depicted with a blue color). Secondly, they incorporated the requirement of moral capabilities of the actors, by depicting them with a big heart (full human or human-like moral capabilities), a small heart (some moral capabilities, but not at the human level), or no heart (indicating no moral capabilities).

| **Concept** | **Type** | | **Example in scenario** | **Depiction** |
|---|---|---|---|---|
| *Actor* | Human | | Jason | 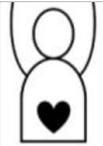 |
| | Machine agent | | Robotic guide dog | 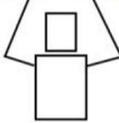 |
| *Work* | Direct | Moral | 'decide to change route' | 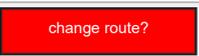 |
| | | Amoral | 'walk', 'guide', 'recognize traffic light' | 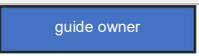 |
| | Indirect | Moral | 'scan for abnormalities' | 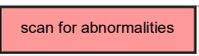 |
| | | Amoral | 'hand over tasks' | 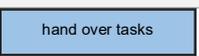 |
| *Moral capacity* | Human | | Agent has human or human-like capacity for recognizing and making moral decisions | 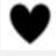 |
| | Partial | | Agent has some capacity for recognizing and making moral decisions, but not at human level |  |
| | None | | Agent has no capacity for recognizing or making moral decision | |

| | | | |
|---|---|---|---|
| *Asymmetry* | | Task division is asymmetric, e.g. 'give support', 'supervise' | |
| *Transition* | Automatic | | |
| | Initiative | If Jason notices an abnormality, he initiates a transition. | |

**Table 1:** Important concepts in the Team Design Pattern language and their graphical depiction. The third column shows examples from the example scenario shown below.

*Example scenario*

As an example, let us look at an example scenario of a blind man called Jason. Recent technological advancements allow the design of a robotic guide dog that can support Jason in his daily needs, together constituting a human-agent team. Designers of the robotic dog can use TDPs to explicate the expected behaviors, requirements, and responsibilities in this (elementary version of a) hybrid intelligent system. Let us imagine the envisioned robotic dog guiding Jason to a doctor's appointment. They are walking on a countryside road and encounter a T-intersection. They need to turn right to reach the doctor's office, but suddenly a cry for help is heard from the left. What should the team do? Figure 1 shows three possible TDPs to describe this situation in increasing levels of complexity.

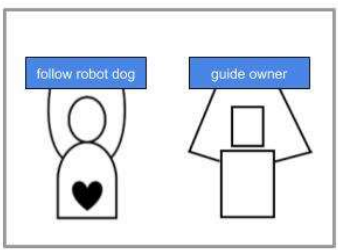

**Example pattern 1:** This pattern shows the basic tasks: Jason follows the robot dog, and the robot dog guides their owner. Just by itself, this example is not capable of describing how the human-agent team should react to the cry for help. Using this pattern in the development of the system could result in the team simply ignoring it *by design*.

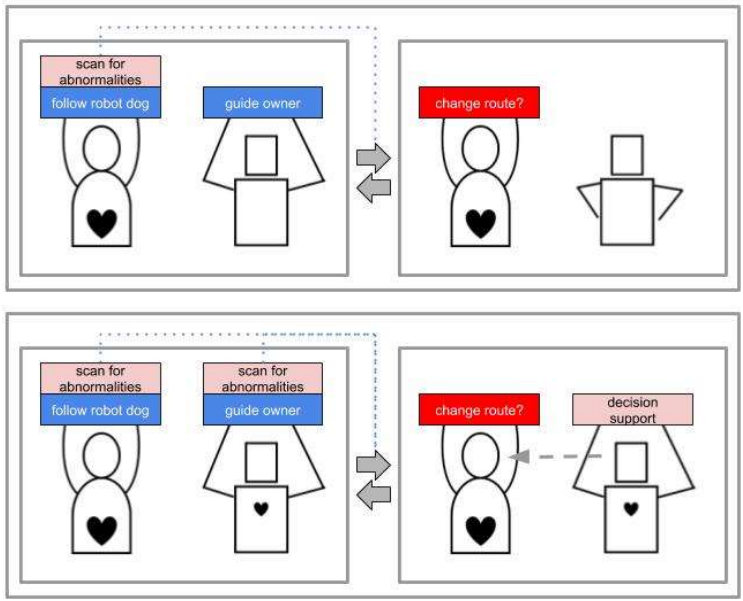

**Example pattern 2:** Jason has another task while following the robot dog: scanning for abnormalities. Since this task does not necessarily bring the team closer to the doctor's office, it is an indirect task. Additionally, because the task may require some moral cognitive components, it has a red color. If Jason finds any abnormalities, he can take the initiative to change the task division, resulting in the transition to a different frame. Now, the robotic dog stands still, while Jason decides to change the route.

**Example pattern 3:** This pattern is similar to pattern 2, but the robot dog has more responsibilities. Not only Jason, but also the robotic dog can scan for abnormalities and initiate a transition. This may be useful because the robotic dog may be able to perceive visual abnormalities or even have better hearing than Jason. After the transition, when Jason is deciding whether to change the route, the robotic dog can provide decision-support. For example, the robotic dog may be able to assess the level of urgency of the person in need or remind Jason of the urgency of his own doctor's appointment.

**Figure 1:** Three patterns describing the example scenario of Jason walking with his robotic dog.

### 2.1.2 Socio Cognitive Engineering and Value Sensitive Design

*Socio Cognitive Engineering*

While Team Design Patterns have received some attention in recent years, scholars have thus far mainly focused on providing examples of what the design patterns could look like. A standardized set of methods for the creation of Team Design Patterns is currently still missing, even though this is a vital component for a successful universal adoption of the design tool across a range of disciplines. This methodology has several requirements. It should (1) be geared to hybrid intelligence by combining technical AI knowledge with human factors knowledge, (2) be able to incrementally improve the patterns based on its application to new use cases, (3) in the case of moral decision-making, be able to incorporate moral values of the stakeholders involved in the pattern's application domain.

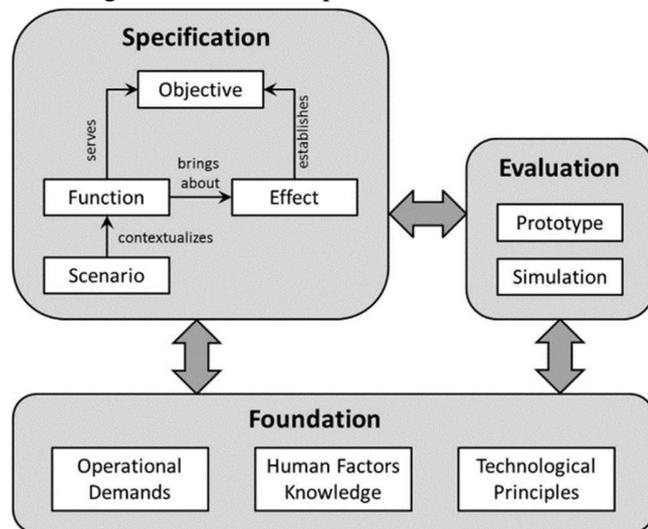

**Figure 2:** An overview of the Socio-Cognitive Engineering method by Neerincx et al. (2019)

Socio Cognitive Engineering (SCE, Neerincx et al., 2019) is a methodology that meets the first two requirements. It was developed specifically for the design of hybrid intelligent systems, combining elements from cognitive engineering, user-centered design, and requirement analysis. In the past decade it has been implemented in a wide range of systems in various domains, including digital assistants supporting children with diabetes (Looije et al., 2016), social robots for the elderly (Peeters et al., 2016), and a support system for disaster response (Mioch et al., 2012). An overview of SCE is illustrated in Figure 2. The methodology is always applied to satisfy a specific demand in a particular context. The *foundation* layer consists of three components. Firstly, it includes an analysis of operational demands, which revolves around inspecting the work domain of the hybrid intelligent system and the support that is needed. Additionally, it includes an analysis of human factors relevant to the system, as well as an analysis of the technological principles that may be appropriate for the envisioned support. In the *specification* component of SCE, a number of objectives of the envisioned system is defined. This leads to the recognition of functions of the system, which are contextualized by scenario-like descriptions of the supposed human-machine interactions. Finally, the system's functions are supposed to bring about certain effects, which are called a claims. Lastly, the *evaluation* component aims to use a prototype or simulation to test whether the specified functions really have the claimed effect in order to fulfill the objectives. The results of the evaluation can then be utilized to revise and enhance the foundation and specification components, incrementally advancing the product. Neerincx et al. (2019) emphasize that SCE can be used in a cyclic fashion, incrementally advancing the foundation, specification, and prototype of the system. However, there are no strict rules in the order of revisions, as the designers can utilize the concepts in the framework to advance their design in whichever order they see fit (Ibid.).

*Value Sensitive Design*

As mentioned before, SCE was not specifically developed for the design of Team Design Patterns, but it is a promising methodology in doing so. To satisfy the third requirement mentioned above, the moral values of the stakeholders of the patterns must be incorporated in the methodology. Value Sensitive Design (VSD) is the most elaborate of methodological frameworks to deal with values in the design process (Harbers & Neerincx, 2017). It defines values as *'what a person or group of people considers important in life'* (Friedman et al., 2013).

Stakeholders can be direct, meaning that they directly interact with the envisioned system, or indirect, meaning that they do not interact with the system but may still be affected by it. When one value comes at the expense of another, this often results in so-called value tensions. Harbers & Neerincx (2017) proposed to integrate VSD methods in the SCE methodology in each of the three components. Most importantly, this would result in a careful analysis of the direct and indirect stakeholders of the product in the *foundation* layer. When sufficient attention is paid to the values of the possible stakeholders, they argue, this will naturally affect the *specification* phase. This can happen implicitly, when they are used in the process of defining the scenarios, objectives, functions, and claims. Additionally, values can be supported in the specification phase explicitly through value stories, when there is a separate objective to support a certain value. Finally, in the *evaluation* phase it is necessary to test the effects of the prototype on the values of the stakeholders.

Harbers & Neerincx (2017) used this combination of SCE and VSD in the development of virtual assistant for workload harmonization in teams of railway traffic control operators in The Netherlands. To discern the operational demands, they organized a workshop resulting in a list of stakeholders and their values. They found that the most relevant values were insight, helping others, and privacy. This attention to values automatically resulted in the objectives in the *specification* component explicitly supporting these values. Finally, the design process led to a prototype virtual assistant that was positively received by the stakeholders and domain experts.

While the SCE and VSD methodologies are not specifically geared towards the design of Team Design Patterns, this chapter has thus far shown their suitability in doing so. I propose treating the Team Design Patterns as a special and abstract prototype that can be created through using the foundation, specialization and evaluation concepts as described above. However, TDPs are special in the sense that their stakeholders are not just the people that will be working with and affected by the envisioned system in its application domain (e.g. the operator, team leader, and passengers in Harbers & Neerincx' study), but also the designers of the system. As these researchers and developers are the primary prospective users of these patterns, it is key to include their wishes and demands in the three phases of the methodology as well.

## 2.2 Research design

The research of the current thesis is structured according to the three phases of the SCE methodological framework. As mentioned in the previous section, this methodology is always used to solve a problem in a particular context. For this thesis, this requires a use case of a system in development in which moral decisions are made. The use case that this thesis focuses on is a hybrid intelligent system aimed at countering Diabetes Type II (DT2) in the Netherlands, being developed by the Netherlands Organization for Applied Scientific Research (TNO). This system will take the shape of a digital assistant for both health care professionals and patients in the prevention, diagnosis, treatment, and management of the disease. The system is hybrid because it requires a close cooperation between human knowledge and social skills on the one hand, and machine-based learning on the other. The DT2 use case is embedded in a four-year research program called FATE, aimed at researching novel techniques for fair, transparent, explainable (co-)learning decision making in human-machine teams. The research presented in this thesis roughly coincided with the first half year of this five-year project, meaning that the research project was in its very early stages. Hence, the envisioned DT2 system was treated as a use case and not (yet) as a system ready for deployment.

The current research is largely exploratory by nature, as it aims to acquire knowledge about ways to conceptualize moral decision-making in medical hybrid intelligent systems in such a way that it may help researchers in communicating about such systems and make them more ethical by design. Both the research questions and the proposed methodology to answer them are dissimilar to any research in this field thus far. Therefore, the aim of the research design is not to acquire definitive answers, but rather to enhance the pattern language and sketch a framework for the methodology for creating patterns in the domain of moral decision-making. This can serve as a starting point for prospective cycles of the SCE methodology and future research in general. The foundation, specification, and evaluation components are used as a guidance for the creation of a first HI pattern language for AI-based decision support systems, particularly geared towards decisions with moral aspects in lifestyle-related health management.

# 3. Foundation
## 3.0 Introduction

The previous chapter described the SCE and VSD methodological frameworks and explained the motivation for adopting them for the research in this thesis. It also presented the general research design, as well as the DT2 use case. The current chapter describes the foundation layer of the SCE methodology as shown in Figure 2, consisting of three components: the analysis of operational demands (section 3.1), relevant human factors (section 3.2), and relevant technological principles (section 3.3).

The analysis of operational demands consists of an investigation into the current DT2 care in the Netherlands, as well as the vision of the anticipated system. This research focuses on what type of support is needed in the envisioned system. The analysis was done in an explorative fashion, drawing information from academic literature, medical guidelines, and team members of the FATE program. Additionally, the inclusion of VSD led to an investigation of the stakeholders and their values in the analysis of the operational demands. This was based on medical guidelines, first drafts of the envisioned system, and an interview with four experts in the domain of lifestyle-related diseases and their care. Finally, this led to the identification of three main design challenges in which moral decision-making has a large role: (1) the mitigation of harmful biases in learning algorithms, (2) the sharing of patients' medical and behavioral data and the recording of consents, and (3) suggestions and interventions into the patients' lifestyle. The former two design problems were taken as use cases for the current research.

The analysis of human factors focuses on the current academic knowledge regarding moral decision-making. The analysis of technological principles is geared towards existing solutions for the two design solutions identified in the analysis of operational demands. This analysis was based on both academic literature and interviews with experts in the domains of bias mitigation and data privacy.

## 3.1 Operational demands

The operational demands analysis of the SCE methodology revolves around the question: What kind of support is needed in the application domain? (Neerincx et al., 2019) In its early design stages, the aim of the envisioned

DT2 system is to provide support to both health care professionals and patients in the prevention, diagnosis, treatment, and management of the disease, leading to a large number of possible applications. However, the general structure of the FATE system is shown in Figure 3. AI developers and healthcare professionals create models based on existing patient data and domain knowledge. These models may, for example, be geared towards predicting patient's risk of developing diabetes (aiding prevention) or suggesting a diagnosis. Alternatively, they may predict the best type and dose of medicine (aiding treatment) or predict the best type of lifestyle change to make the disease as unintrusive as possible (aiding management). Through several modules, these predictions or suggestions are presented to the relevant patient or healthcare professional. Finally, the models are improved and updated by patients' medical and behavioral data.

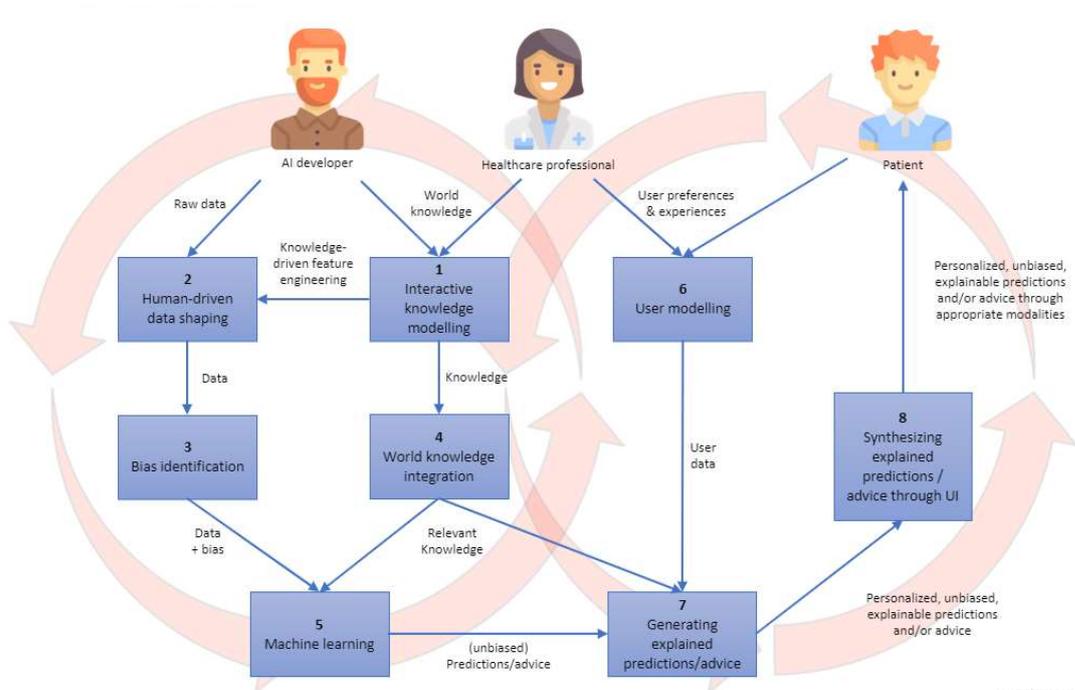

**Figure 3:** An overview of the general structure of the anticipated DT2 system.

As the current thesis focuses on the support of moral decision-making, the next step in this analysis is to localize processes in this system in which actors face choices that may have a moral component. The VSD method of listing all the possible stakeholders and their values was used. The direct and indirect stakeholders and their most important values were extracted from interviews with medical experts and designers of medical HI systems. The results are listed in Table 2. This list may not be exhaustive but can be complemented in future iterations of the SCE method. It is likely that value-tensions in the envisioned system overlap with processes where moral decision-making occurs, as a choice is often seen as having a moral component when the choice or its effects constitute a trade-off between values (Dignum, 2019).

The general structure of the DT2 system and the overview of stakeholders and their values was used as the basis for a semi-structured interview with four experts in the field of lifestyle medicine and digital health. Together they identified three morally sensitive processes in the prospective system, where value-tensions may arise (shown in Table 3). The current thesis focuses on the former two challenges (bias mitigation and data sharing), because of time constraints and because they are the most tangible (the third problem only arises in later stages of the DT2 system's development).

| Stakeholder | Values |
|---|---|
| **Patient** | Health |
|  | Quality of life |
|  | Autonomy |
|  | Habits |
|  | Appearance |
|  | Social life |
|  | Privacy |
|  | Safety |
| **Family member** | Health |
|  | Quality of life |
| **Healthcare professional** | Beneficence |
|  | Non-maleficence |
|  | Patient's autonomy |
|  | Justice |
|  | Efficiency |
| **AI developer** | Health |
|  | Quality of life |
|  | Effectivity of system |
|  | Satisfaction of shareholders |
| **Government** | Health |
|  | Quality of life |
|  | Safety |
|  | Ability to work |
|  | Durable healthcare system |

**Table 2:** VSD analysis of direct and indirect stakeholders of the envisioned DT2 system.

*Bias mitigation*

The first moral problem is that learning models may develop biases that may result in unfair treatment by the system. The underrepresentation of certain ethnicities in the input data, for example, can result in a racial bias in the system. People of those ethnical backgrounds may then receive worse care than others, which can be unfair and unjust. The system's developers may employ techniques to mitigate the harmful bias (see section 3.3.2), but this usually results in a lower average accuracy of the system. The direct stakeholders in this moral issue are the AI developer and the patients. The most prominent value-tension is between the system's (average) effectivity and fair treatment of each patient.

| Process | Stakeholders | Value-tensions |
|---|---|---|
| Bias mitigation of learning models | AI developer | System's effectivity |
| | Patients | vs. |
| | | Fairness |
| Sharing of health and behavioral data | Healthcare professional | Beneficence |
| | AI developer | vs. |
| | Patient | Privacy |
| Lifestyle interventions | Healthcare professional | Beneficence |
| | AI developer | vs. |
| | Patient | Autonomy |

**Table 3:** Overview of possible value-tensions in the prospected DT2 system. Due to time constraints, the current thesis focuses on the former two processes.

*Sharing of health and behavioral data*

The second moral problem that may arise in the prospected DT2 system is the sharing of patients' health and behavioral data. This is useful for the healthcare professionals because they have an increased ability to track the development of the disease in their patients, allowing more personalized care and quicker responses to urgent situations. The tracking and sharing of patient data also results in improvements in the models of the system, as these may be self-learning when they get the true values of their predictions. Additionally, they can detect patterns in large amounts of health and behavioral data, resulting in more personalized care. However, the designers of the system have a moral and legal obligation to ask for consent to share the patient's data to their doctor or to use it to enhance the system's models. The data is the patients' property and their autonomy may be compromised when it is shared without their consent. Yet, although it is key to respect the patient's autonomy regarding their own data, some patients may find it confronting, distressing, or simply too complex to do their research in how and to whom their data will be shared. The direct stakeholders in this moral issue are the systems' designers, the healthcare professionals, and the patients. The most predominant value-tension is between beneficence (resulting from more personalized care from the healthcare professional and the system) and the patient's privacy.

*Lifestyle interventions*

In the treatment and management applications of the envisioned system, the learning models are expected to give suggestions for altering the patient's lifestyle (e.g. more exercise, less smoking, a different diet), possibly based on the individual's unique profile. Additionally, the possibility to track the progress of the disease based on the patient's (bio)medical and behavioral data may result in better and more personal suggestions from health care professionals. These suggestions would likely be visible on a mobile device, in the shape of an eHealth application. This functionality has unprecedented potential for providing personalized care but has a risk of being experienced as invasive. The direct stakeholders in this moral issue are the systems' designers, the healthcare professionals, and the patients. The main value-tension in this moral problem is between beneficence (due to more personalized care) and the patient's autonomy.

## 3.2 Human factors

In very simple terms, morality is a feature of an action or a person that can be considered to be on a scale from good to bad. Moral decision-making has received substantial attention for as long as philosophers have existed, as questions about what is "good" and how one can achieve it have been at the center of many a debate. Throughout history, research into morality can be divided into two very high-level categories: prescriptive and descriptive. The former category, also called ethics, has received most attention, and focuses on answering questions such as what is good and how one *should* act to arrive there. The latter category is concerned with studying how people actually make decisions that fall in the moral domain and has only developed with the rise of academic disciplines such as psychology and econometrics. It is often difficult to keep these categories entirely separated, as the descriptive study of decisions that fall in the moral domain inherently necessitates the delineation of what counts as good or bad, which is a normative presupposition.

Nevertheless, the normative-descriptive distinction is helpful to structure the description of academic theories in this domain. Hence, the following subsection will touch on the wide debate regarding normative morality, or ethics. The second subsection will then give an overview of descriptive accounts and models regarding moral decision-making. The latter section is most directly relevant for the current research, as this study attempts to describe and conceptualize *actual* moral decision-making in human-computer teams. However, the current thesis is not merely descriptive in nature but also attempts to *design* optimal structures for moral decision-making, which can facilitate and promote ethical behavior of the human-computer team in question (Hancock, 2003). Hence, the definition of what is *successful* moral decision-making is an inherently ethical question.

### 3.2.1 Ethics

In broad terms there are three major approaches in normative ethics: (1) virtue ethics, (2) consequentialism, and (3) deontology. Virtue ethics, originating in the work of Aristotle, revolves around the development of a moral character. It treats virtues and vices as the foundational components of morality, meaning that achieving 'good' is by developing virtuous character traits (Hursthouse, 1999). As the name suggests, consequentialism treats the consequences of actions as the foundational components for ethical behavior. The best way to do 'good', according to this theory, is trying to predict the consequences of your optional actions and pick those that maximize well-being (Sinnott-Armstrong, 2003).

Finally, deontology (or duty ethics) emphasizes moral rules as the basic component of morality. According to this theory, initiated by Immanuel Kant, the most ethical way to behave is to act in accordance with a set of universal rules that translate into rights and duties. An important manifestation of this approach is visible in the United Nations' Universal Declaration of Human Rights (UN General Assembly, 1948). In bioethics, an important and influential subset of duty ethics is Beauchamp and Childress' widely accepted four principles of bioethics (2001). The set of principles is made up of respect for autonomy, beneficence, non-maleficence, and justice, which can largely be traced back to the ancient Greek Hippocratic tradition.

### 3.2.2 Descriptive accounts of moral decision-making

While the philosophical debate about how one *should* act is still very alive, the rise of the field of psychology has paved the way for a different scientific angle towards morality: the descriptive account of how people *actually* take moral decisions (regardless of whether they pick the ethically "right" choice).

*Developmental models*

One of the most influential theories in moral psychology was coined by Lawrence Kohlberg (1958). He described the development of moral reasoning in humans as a transgression through multiple stages, inspired by Piaget's stage theory of cognitive development (1932). Based on empirical research on children, he concluded that humans necessarily transgress from 'pre-conventional' stages of reasoning (based on egoism, obedience and punishment) to 'conventional' stages of reasoning (based on social norms or law and order), and ultimately to 'post-conventional' stages of reasoning (based on social contracts and universal ethical principles). As in Piaget's work, stages cannot be skipped because they each provide a new perspective in the development of reasoning that is necessary for the following stage (Dovidio et al., 2017).

While Kohlberg has used a body of evidence to support his step-wise developmental model of moral reasoning, the field of behavioral psychology soon found that moral reasoning as specified in Kohlberg's theory does not account for people's actual moral behaviors (Ibid.). In fact, there is a general academic consensus that there is a considerable gap between people's moral ideas and putting those to practice. This was famously revealed by Darley & Batson (1973), who showed that people who were in a rush to give a lecture on the topic of 'Good Samaritans' were unlikely to offer help to a person in need. Several lines of research have attempted to explain this gap between moral reasoning and moral behavior. Sternberg (2012), for example, developed a chronological psychological model of 8 steps that need to be taken for successful moral decision-making, as shown below. If the subject fails to take any of the 8 steps, they will not engage in ethical behavior (regardless of their conscious moral principles).

1. recognize that there is an event to which to react;
2. define the event as having an ethical dimension;
3. decide that the ethical dimension is of sufficient significance to merit an ethics-guided response;
4. take responsibility for generating an ethical solution to the problem;
5. figure out what abstract ethical rule(s) might apply to the problem;
6. decide how these abstract ethical rules actually apply to the problem so as to suggest a concrete solution;
7. prepare for possible repercussions of having acted in what one considers an ethical manner;
8. act.

In Sternberg's view, his model is not only an accurate depiction of successful moral decision-making but can also be used for the "moral education" of kids and subsequently close the gap between theory and practice. His perspective is strongly embedded in western rationalism, as he explicitly assumes that ethical reasoning and ethical behavior can be largely rational (Ibid). This is a controversial notion, which I will soon return to. However, Sternberg points out two important requirements for moral reasoning that are relevant for any model. Firstly, he emphasizes that the person in question needs to '*define the situation as having an ethical dimension'*. For any choice in any situation, the interpretation of that choice as an *ethical* one has a profound impact on how one reacts to it. This component provides one answer to the perceived gap between people's ethical ideals and their practice: they often simply do not recognize the situation in which their (or any) moral values apply. If a problem is not perceived or defined as a moral problem, someone will most likely also view their reaction and its consequences as amoral. Secondly, in the fourth step, Sternberg stresses that taking *'responsibility for generating an ethical solution to the problem'* is an essential requirement for engaging in ethical behavior. Many would attest that a large factor in most of the world's largest problems is the fact that many recognize them, but few feel responsible for their solutions. Especially in organizational structures where it is not immediately clear who bears responsibility for the ethical outcome of certain problems, it can thus be advantageous to make explicit agreements in order to ensure that someone is accountable for reacting to a moral decision.

*Moral intuitionalism*

However, these models trying to explain moral reasoning have received widespread criticism on being too rationalistic. Dennis Krebs and colleagues (1997) noted that till that moment, moral decision-making had nearly solely been studied using abstract moral dilemmas, while real-life moral decision-making has many other shapes that are much less philosophical and more social in nature. Around the turn of the millennium, Jonathan Haidt (2001) posed the hypothesis that not moral reasoning, but moral emotions and intuitions are the origin of moral judgement and decision-making. In his view, people perform moral reasoning in a so-called 'post-hoc rationalization' only to justify moral judgements that have already been formed by immediate moral intuitions (Ibid). This 'social intuitionalist' model is supported by compelling evidence and soon gained much attention, not the least from evolutionary psychologists. The idea that emotions play a large role in moral judgement is sensible from an evolutionary perspective, as most theorists agree that morality has developed in humans (and other species) as a mechanism to ensure cooperative and pro-social behaviors in groups (Tomasello & Vaish, 2013). A group or society in which individuals hurt each other and play by their own rules is probably less successful and has a lower chance of survival than a group that shows cooperative behavior towards one another. In the words of Dennis Krebs and colleagues, the original function of moral judgement was 'to induce those with whom one formed cooperative relations to uphold the cooperative systems in order to maximise the benefits to all' (1997). Subsequentially, groups with 'moral' behavior had a higher chance of survival, leading to their genes (and ideas) flourishing, which in turn led to both a natural and social ingraining of moral responses to certain situations. As the spread of these attitudes through genetics nor social interaction is necessarily dependent on language, which makes it possible for them to be perceived as 'intuitions' (and may explain why they are sometimes so hard to put to words). This evolutionary function fits the development of social emotions such as shame, anger, envy, and guilt.

Yet the intuitionalist perspective on morality has its own downsides, as it alludes to the idea that morality is ruled by emotions and cannot be influenced by conscious thought. This begs the question whether we have any control over our arguably most important decisions, while at the same time discounting an entire body of rationalistic literature.

*Dual-process theory*

The rationalist and intuitionalist views on moral decision-making were united in dual-process theory. This theory encompasses the idea that the functioning of the brain can be divided into two 'systems' or 'pathways'. One of the pathways is 'fast', evolved early in the development of humans and animals, and is often associated with quick, automatic, and emotional responses. The other pathway is 'slow', ends in the prefrontal cortex, and is usually associated with conscious and controlled thought (Sloman, 2002). The notion of these two brain mechanisms explaining human behavior has been deeply influential for the past two decades, affecting virtually every subdiscipline in psychology, neuroscience, and beyond. In a persuasive series of articles (2004, 2007, 2008, 2009), Joshua Greene and colleagues applied dual process theory to moral reasoning in an attempt to synthesize the rationalist and intuitionalist models described above. The basis of their proposed theory can be found in neuroscientific fMRI research in 2001 (Greene et al., 2001). In this study, participants faced traditional 'trolley problem' dilemmas (Thomson, 1984). The brain imaging indicated that deontological moral judgement is associated with neural activity in the fast, emotional system in dual process theory, while utilitarian moral judgements go together with activity in the slower, conscious system. While the contraposition of two competing driving forces for human behavior – one rational, one emotional – is an old and familiar picture, aligning this distinction with the debate between normative moral theories and mapping these on brain processes was a ground-breaking enterprise.

Greene's theory would not be influential without being the subject of sobering criticism. In response to dual process theory in general, many authors have stressed that the high-level distinction between two systems does not do justice to the complex and interactive nature of the brain (e.g. Osman, 2004; Pennycook et al., 2018). While it may seem attractive to classify all human thoughts and behaviors as resulting from a one-dimensional distinction in brain processes, this reduces the perceived complexity of the processes, eventually making it harder to understand them in their entirety. Additionally, there is always a risk of identifying the competition between the two systems as the cause of reasoning and behavior, while the evidence is merely of correlational nature. Finally, Greene's experiments have been subjected to a long list of methodological objections (Berker, 2009). For example, scholars have noted that the moral dilemmas used in the research can better be described as 'personal versus impersonal' than 'typical retributivist versus typical consequentialist' (Ibid.). This may explain the neuroscientific and behavioral differences found in their study better than a distinction between two philosophical ideas. This supports the wider criticism that it is unlikely that a twofold distinction in brain processes maps perfectly on an ancient and abstract philosophical debate.

Regardless of these denunciations, it is important to take Greene's dual process theory seriously when researching human moral reasoning and behavior. People's moral decision-making can (at least partly) be explained as an interplay between (sometimes competing) mechanisms that also determine behavior in non-moral domains (Bucciarelli et al., 2008). Sloman's dual process theory (2002) may serve as a first tool for structuring further research.

*Discrete moral choice model*

The study of people's reaction to moral choices has not solely received attention in psychology and cognitive neuroscience. Recently, economists and econometrists too have entertained the topic. Caspar Chorus (2015), for example, applied the economic perspective of discrete choice analysis in an attempt to synthesize many of the models mentioned above. Figure 4 shows the resulting model. Where Greene's model was criticized for lumping all human morality in two big categories, Chorus shows that moral behavior is a subtle interplay between several long- and short-term processes. This includes feedback-loops of behaviors, expectations, and changing personal and societal norms due to the behavior of the individual and others.

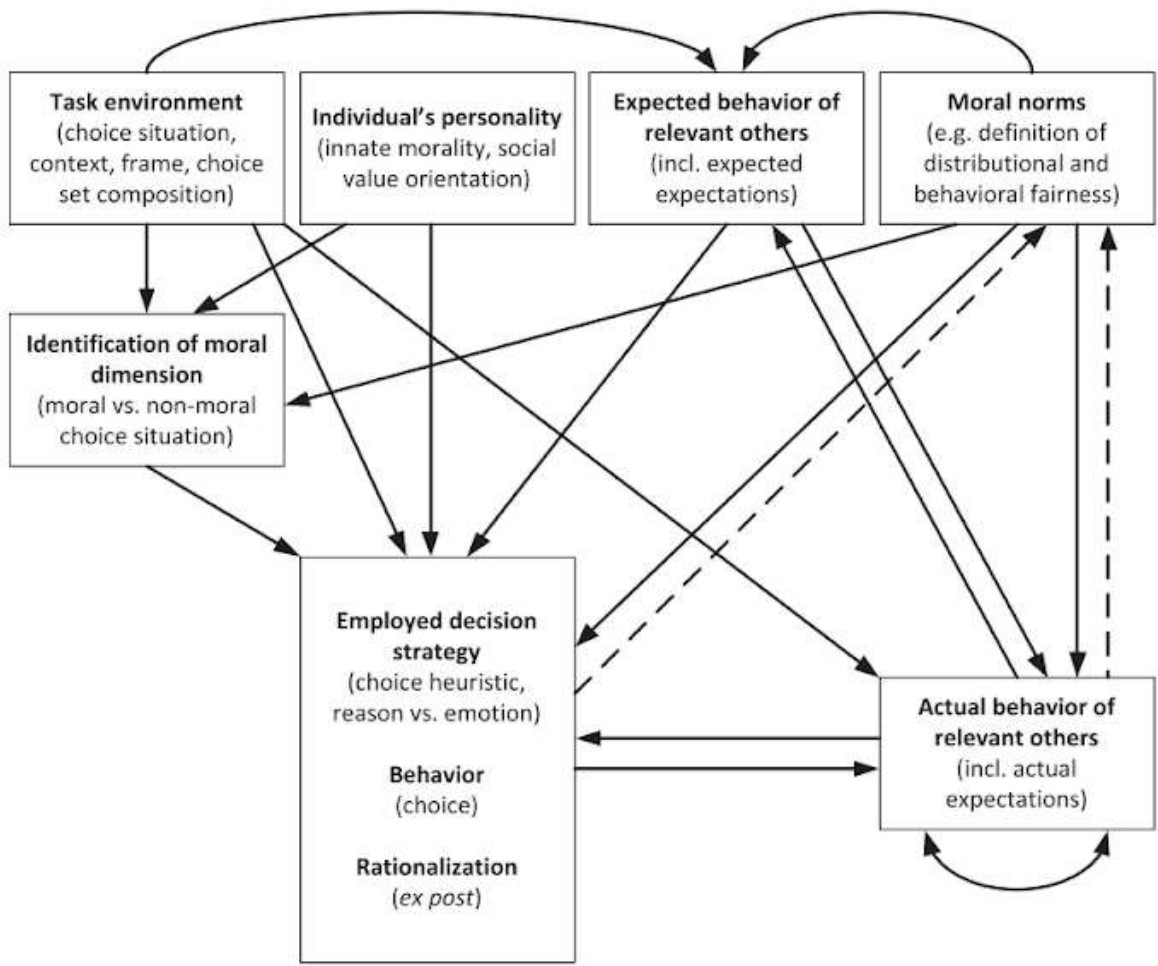

**Figure 4:** The conceptual model of an individual's moral choice behavior as described by Chorus (2015). Arrows are short-run causal impacts, dashed arrows are long-term feedback loops.

Chorus stresses that his intricate figure only shows that *'the nature and (social) origins of moral decision making are far too complicated and subtle to incorporate into one choice model or data collection effort'* (2015). Instead, he attests that his model is a first step in deconstructing the inner nuances in moral reasoning and behavior, and that it can be used as a framework for further research.

*Team-level factors*

Finally, research into moral decision-making on a team level is highly relevant to the conceptualization of moral decision-making in human-machine teams. Unfortunately, few scholars have investigated this matter, with a few exceptions. Van Soeren & Miles (2003) examined the role of teams on moral distress in end-of-life decision-making in critical care through a case study. They found that tensions between the stakeholders (including the family, intensive care professionals, and transplant team members) often arose from a lack of communication and feelings of being unheard. They suggest shaping a process of regular interdisciplinary team reviews for all stakeholders to give input from their perspective (Ibid.). According to the authors, this would allow for everyone involved to take a step back from dealing with consecutive crises, and instead look at the overall continuity of the patient's condition. They add that this would benefit a mutual development of trust among team members, resulting in fewer moral tensions in the decision-making process.

In a similar vein, Gunia et al. (2012) researched the role of contemplation, conversation, and explanation in moral decision-making. In conversation style, they differentiated between conversation partners that promoted more self-interested behavior (self-interested conversation) and conversation partners that stimulated more community-minded behavior (moral conversation). They found that contemplation and moral conversation promote more selfless behavior, while immediate choice and self-interested conversation cause more selfish behavior (Ibid.). Additionally, they found that people can usually provide explanations consistent with their decisions just before and just after those decisions.

While it may be unsurprising that these two studies on team processes in moral decision-making emphasize the role of communication processes, it is important to explicate and specify these factors in an attempt to get a clearer overview of how team processes precisely affect decisions in the moral domain.

### 3.2.3 Types of cognitive aids

At the intersection of human factors and technology lies the categorization of different types of cognitive aids. McLaughlin et al. (2019) recently made a useful contribution to this field by linking advancements in aid development to psychological theories of cognition. The authors affirm that traditional taxonomies of cognitive aids focus on their surface characteristics, while at the most fundamental level they can be categorized by which cognitive process they support. McLaughlin and colleagues identify five main cognitive processes: attention, memory, perception, decision-making, and knowledge (Ibid.). Each of these classes consists of several more specific processes. For example, attention aids can support humans in selective, orienting, sustained, or divided attention. The entire taxonomy is shown in Figure 5.

| Attention aids | Memory aids | Perception aids | Decision aids | Knowledge aids |
|---|---|---|---|---|
| • Selective<br>• Orienting<br>• Sustained<br>• Divided | • Prospective<br>• Recognition<br>• Recall<br>• Short-term<br>• Sensory | • Vision<br>• Auditory<br>• Vestibular<br>• Touch<br>• Olfaction | • Individual<br>• Team<br>• Debiasing<br>• Reasoning | • Mental models<br>• Team roles<br>• Creativity<br>• Language translation |

**Figure 5:** A classification of cognitive aids by McLaughlin et al. (2019).

McLaughlin and colleagues acknowledge that this taxonomy is not set in stone, as it is based on current psychological theories and constructs (Ibid.). Additionally, some aids may be difficult to classify as their exact support is ambiguous, or an aid may provide several types of support at the same time. However, this basic-level classification may help in the construction of Team Design Patterns in order to specify which type of support is required in a certain situation. As the previous subsection demonstrated, moral decision-making is a complex mental and social process for which philosophers and scientists do not have a univocal explanation. It is apparent, however, that the process requires all five categories specified by McLaughlin et al. (2019). The current thesis aims to use these five main cognitive processes to conceptualize technological requirements in the Team Design Pattern language.

## 3.3 Technological principles

As mentioned in chapter 2, the envisioned system to support the prevention, diagnosis, treatment, and management of DT2 is still in its very early stages at the time of writing this thesis. The necessary technologies in terms of hardware and software are still largely unclear. The general vision of the system is that it will be able to provide personalized care on a range of (mobile) devices, using modules that combine different types of artificial intelligence. Subsection 3.3.1 gives a short overview of the high-level distinction between machine learning and knowledge reasoning systems. The subsequent two sections give an overview of possible solutions the academic literature prescribes for the bias mitigation and data sharing design problems.

### 3.3.1 Hybrid AI

The envisioned DT2 support system is expected to make use of artificial intelligence to provide personalized care. The current literature differentiates between two main types of artificial intelligence (Van Harmelen & Ten Teije, 2019). On the one hand, there are data-driven machine learning techniques, commonly used for identifying rules or patterns in large amounts of data. On the other hand, there are knowledge reasoning techniques, usually model-based and useful for deductive practices including existing world knowledge. Until recently, the academic community studied these two classes of AI separately, but in recent years the attention has shifted to the combination and integration of the two types of methods (e.g. Van Harmelen & Ten Teije, 2019; Dijk, Schutte & Oggero, 2019). This is highly relevant for the DT2 support system. For example, a classical diagnosis task normally consists of mapping existing medical knowledge onto a model and using a knowledge reasoner to navigate the model in order to classify a new patient. However, the use of large amounts of patient's medical and behavioral data can lead to the discovery of correlations or patterns that are not even known to the medical world yet. After discovery, these new rules may be added to the knowledge reasoning model, improving the quality of diagnoses.

   Van Harmelen & Ten Teije (2019) have created a pattern language as a taxonomy to classify the inner mechanisms of hybrid systems on a high level. While it is outside the scope of this thesis to fully integrate these 'boxology design patterns' with the Team Design Pattern language, the high-level differentiation between machine learning methods and knowledge reasoning techniques (and their combinations) are highly relevant to the technological conceptualization of hybrid intelligent systems.

### 3.3.2 Bias mitigation technologies

Machine learning algorithm fairness has received considerable attention in recent years, resulting in a plethora of different definitions of fairness and bias. Machine ethicists have studied what is meant when we want people to be treated fairly and have defined at least two conditions: group fairness and individual fairness (Zemel et al. 2013). Group fairness indicates that the proportion of subjects in a protected group being classified positively is equal to the proportion of the population as a whole. In other words, the chance of receiving a false positive or false negative should be independent from certain sensitive features (e.g. gender, race, sexuality, religion, etc.). This is also known as statistical parity (Ibid.). Individual fairness, in contrast, entails that people with similar traits with respect to a certain task be treated similarly. While these are intuitive conditions, it has proven difficult to find a universal measure to assess whether an algorithm meets these conditions. Till Speicher and colleagues (2018) proposed the use of existing inequality indices from the field of economics to assess the degree of fairness of algorithms, addressing both group and individual fairness.

In a comprehensive review of the currently available methods for reducing unfairness in machine learning algorithms, Friedler et al. (2019) identify three types of methods based on how they affect the algorithm: pre-processing, in-processing, and post-processing methods. Firstly, pre-processing methods have recognized that training data is a frequent cause of unfairness, as it can capture historical discrimination or under-represent minority groups. Calmon and colleagues (2017) provided a probabilistic framework for discrimination-preventing preprocessing. They defined an optimization problem for dataset transformations that trade off group fairness, individual fairness, and overall accuracy (defined as data utility), using regular classifiers.

In-processing methods consist of modifications to existing algorithms in order to reduce unfair predictions. Kamishima and colleagues (2012), for example, attempted to create a 'prejudice remover regularizer' that enforces a classifier's independence from sensitive information. The method was applied to a logistic regression classifier but can be adapted to be geared to different types of algorithms.

A third fairness-enhancing approach is to modify the results of a classifier through post-processing methods. For example, Kamiran et al. (2012) created a method to modify decision tree leaf labels after training. Pleiss et al. (2017), too, address simple post-processing methods to satisfy fairness constraints. However, they critically note that these methods often hinge upon withholding predictive information for randomly chosen inputs (Ibid.). Many would agree that this is unsatisfactory in sensitive settings such as healthcare, as it implies that individuals for whom a correct prediction or diagnosis is available purposefully receive a false prediction for the sake of achieving equal accuracy for all subgroups.

As is clear from this brief overview, methods for addressing bias and unfairness in learning algorithms are still in rapid development and are likely to change substantially over the coming years. Moreover, it is evident that the suitability of fairness-enhancing measures and methods are dependent on the input data, learning algorithm, and development aim at hand. An HI bias mitigation framework should allow for rapid developments in the field and for the possibility to test and apply methods on a case-to-case basis.

### 3.3.3 Data sharing technologies

The sharing of medical data is a widespread phenomenon, especially due to the increasing popularity of eHealth applications (Blenner et al., 2016). In recent years, several legal requirements have been put in place to safeguard user privacy in digital environments with a strong emphasis on consent (e.g. General Data Protection Regulation (GDPR), 2018). In a comprehensive study of eHealth software for diabetes care, Blenner et al. (2016) found that although many of these applications have privacy policies that formally fulfill the legal requirements, they can be misleading for their users. Patients may, for example, have the mistaken belief that their health data is not shared with third parties, even though this is generally the case. While the scientific realm has formulated ethical guidelines for the data sharing and consent-giving process in smart health (e.g. Jones & Moffitt, 2016; O'Connor et al., 2017), this has not yet been converted into a concrete set of best practices for the design of eHealth systems.

However, several scholars have provided typologies of consent in (digital) medical settings, which can provide a first framework for the technological underpinnings of its design. Firstly, the most elementary form of medical consent was coined *simple consent*, relevant for minor medical procedures that pose a low health risk (Whitney et al., 2003). The simple consent procedure consists of a short explanation of what the intervention entails, followed by an explicit or implicit agreement by the patient. Conversely, recent years have witnessed considerable attention for the concept of *informed consent*: the requirement to ensure that the patient truly understands what they are consenting to (Grady, 2015). According to Whitney and colleagues, the process of achieving informed consent should consist of a 'discussion of nature, purpose, risks and benefits of proposed intervention, any alternatives, and no treatment, followed by explicit patient agreement or refusal' (2003). However, Christine Grady (2015) observed that the informed consent process usually differs substantially in detail and formality depending on whether it is intended for clinical interventions or for research purposes. She also noted that 'with more recently embraced learning paradigms, these goals are converging, or at least the boundaries are shifting' (Ibid.), calling for a more comprehensive and universal set of requirements for the clinical and research consent-giving process.

Sarah Moore and colleagues (2017) recognized the increasing role of digital systems in obtaining consent, making a distinction between *in-person* and *remote consent*. While in-person consent has an inherent social mechanism for establishing mutual understanding and obtaining informed consent, remote consent is less personal, making it more difficult to ensure that the subject has fully understood the terms. In a similar vein, Rowbotham and colleagues (2013) noticed that the dense texts in many remote consent forms resulted in a low degree of the users' attention. They conducted an experiment in which one group of subjects was given a 'standard' consent form, while another group was presented with an introductory video, standard consent language, and an interactive quiz with special attention to data privacy, aggregation, and sharing. The second group had a significantly and substantially better understanding of the research risks and procedures than the control group. This *interactive informed consent* is especially relevant for remote variants of obtaining consent.

The recognition of individual preferences regarding data sharing and the increasing complexity of digital systems has led to the identification of two additional types of consent. Firstly, Rake and colleagues (2017) advocated for a *personalized consent* flow, that would allow subjects to control the health data collected by their mobile and wearable devices, in order to regulate to what extent these are shared for research purposes. This type of consent necessitates a differential set of consent rules that can be divergent for each subject, as well as a mechanism through which the patient can oversee and change these whenever they want (for the latest developments in this domain, see Rau et al., 2020). Finally, Bunnik and colleagues (2013), too, recognized a need for personalized consent rules, but they also noticed that some patients desire a larger choice space (and larger amounts of information) than others. Hence, in a consent system for deciding which hereditary diseases to test for in a personal genome test, they developed a *tiered-layered-staged* model for patients to receive differing amounts of choices and support based on their preference (Ibid.).

# 4. Specification
## 4.0 Introduction

The previous chapter explained the foundation for the creation of Team Design Patterns for moral decision-making in hybrid intelligent systems in the medical domain. This included an operational demands analysis, resulting in three areas in the DT2 system where moral decision-making plays an important role, of which bias mitigation and data sharing were chosen to investigate in this thesis. It also gave an overview of the academic knowledge regarding moral decision-making and current solutions for these two design problems.

The current chapter presents the specification phase of the SCE method, in which all this foundational knowledge is brought together into Team Design Patterns describing possible solutions for the design problems. The operational demands and technology analyses were used to create the patterns' team process configurations. Additionally, the (dis)advantages of each pattern address the value tensions identified in the analysis of operational demands The taxonomy of cognitive aids and human factor literature was used for the patterns' human requirements, while Van Harmelen & Ten Teije's hybrid AI boxology (2019) was used for to identify AI requirements. The patterns' pictures were created using the Google Draw tool.

Section 4.1 addresses the objectives, functions, and claims of the patterns. Subsequently, subsection 4.2.1 presents the patterns for the bias mitigation solution. In the creation of Pattern 2 for this design problem it became evident that another layer of more specific patterns could be created to elucidate one abstract task. These more specific patterns are addressed in subsection 4.2.2. Then, subsection 4.2.3 presents the design patterns for the data-sharing design problem.

## 4.1 Objectives, functions, and claims

*Objectives*

The envisioned Team Design Patterns have three main objectives. Firstly, they are aimed at facilitating communication among researchers and designers from the numerous disciplines that are involved in the creation of hybrid intelligent systems. This may be more complicated than it seems, as it means that the patterns have to cover ethical concepts that are apprehensible for AI engineers and address technical approaches in software engineering in such a way that human factors experts understand it. Above all, these wide ranges of knowledge must be incorporated without being too abstract to provide meaningful conceptualizations of the design problems and solutions.

A second objective of the TDPs is that they need to provide solutions not to one specific design challenge in a single system. Rather, they must be applicable to a set of design problems in various hybrid intelligent systems, preferably in several application domains. This way, the value of the TDPs lies in the possibility to reuse and improve them, drawing lessons from their implementations so far.

Lastly, the use and application of the TDPs should have a positive effect on the envisioned systems. In the case of these patterns aimed at morality in hybrid intelligence, this objective means that the patterns should lead to more thoughtful and explicit moral decision-making than without their use.

*Functions and claims*

The created TDPs have a number of functions that aim to contribute to fulfilling the abovementioned objectives through claims. The most important function of the patterns is their composition: meaning the combination of a textual introduction, a pictorial and textual stepwise representation of the proposed solution, and a table indicating key features of the pattern. Each of these components can be subdivided in smaller functions. For example, each concept of the TDP pictorial language illustrated in section 2.1.1 serves as a requirement for conveying key information of the patterns. Additionally, the tables include the human and AI requirements of each pattern, as well as possible advantages and disadvantages of its implementation. Each of these features can again be understood as functions of their own, as they structure the patterns use and functioning.

The claims of the TDPs correspond to the objectives stated earlier. To facilitate communication between experts from various backgrounds, the claim of the patterns is that they are understandable for designers and researchers across various relevant disciplines. Specifically, the prospective users should understand the solution presented by each pattern (i.e. their content), as well as how the pattern works (i.e. how the functional components relate to and complement one another). Hence, the understandability claim can be subdivided in two claims: *understandability* and *coherency*. These two features may not fully constitute the objective of facilitating communication between disciplines, but they are undoubtedly a fundamental requirement for it. In order to be applicable to more than one situation, the second claim is that the patterns are *generalizable*. The functions should all contribute to a level of abstractness through which the patterns describe solutions that can be reused. Finally, the third claim of the patterns is that they are *effective*, meaning that they lead to better or more appropriate moral decision-making when they are used in the design of a system. There may be a certain degree of subjectiveness in the effectiveness claim, as there is no clear universal notion of the 'appropriateness' of moral decision-making, as this is a normative action. Even though the effectivity claim may be partly subjective, it is still important to the assessment of TDPs as an intended improvement and explication of moral decision-making in hybrid intelligent systems is one of their core objectives.

## 4.2 Team Design Patterns

### 4.2.1 Design Challenge 1: Bias Mitigation

In the envisioned DT2 system, machine learning models will be used to make predictions for patients, for example to give them a diagnosis of pre-diabetes. For Diabetes Type II, it is known that the disease runs a different course in people with a Surinamese or Hindu background than for individuals with European heritage. This means that these different groups also have differing importance of factors for diagnosis (e.g. blood glucose, weight...). Simply removing "sensitive" data types, such as ethnicity, gender, sexual orientation or socioeconomic status may do more harm than good, as they are needed for accurate predictions and help to give the patient the customized care they need. However, the inclusion of these sensitive characteristics while minorities are underrepresented can result in social discrimination: It may bring about systematic differences in the accuracy of the predictions for the minority groups and a misfit of the resulting care to the personal circumstances. As the models are self-learning, the risk of social discrimination remains after deployment, necessitating a mechanism that mitigates harmful biases.

*Pattern 1: Human Decision-Maker*

In this pattern, the machine agent solely performs a classic machine learning task: predicting the diagnosis (e.g. diabetes) of patients. The human AI developer supervises this process, measuring both the overall accuracy and the fairness of the predictions. If the human agent thinks that the balance between these two measures is off (e.g. because people with a Surinamese background receive significantly less accurate diabetes diagnoses), the human AI developer initiates a takeover. In this takeover, the machine stops its task, while the human AI developer changes the model. After he finishes, the situation goes back to normal.

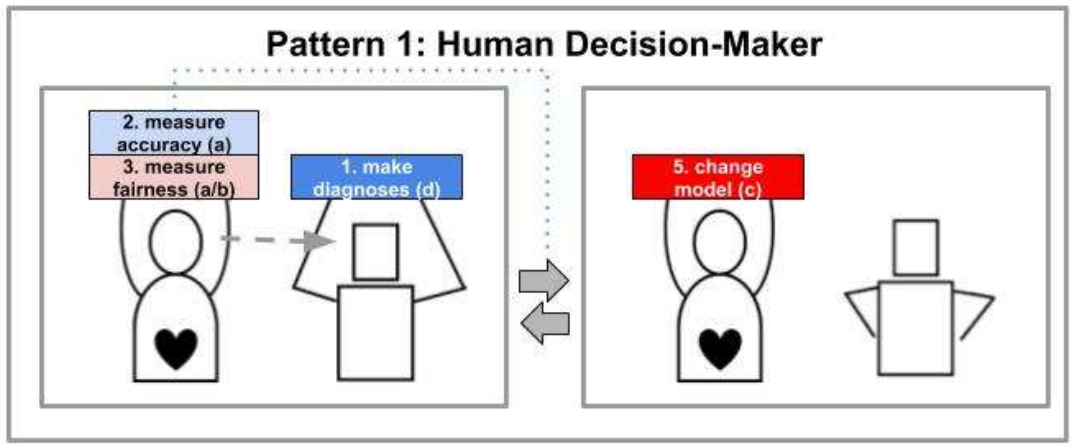

1. Machine agent uses machine learning models to make predictions about diagnosis/treatment
2. Human AI developer performs task supervision: does the model work? Are the predictions accurate?
3. Human AI developer performs moral supervision: are the predictions accurate for everybody? Is there social discrimination based on subgroups?
4. If the balance between overall accuracy and fairness of the model is off, the human AI developer initiates a takeover
5. The human AI developer decides whether to change the model, and how

| Name | 1. Human decision-maker |
|---|---|
| *Human requirements* | a. Sufficient working memory to perform task supervision and moral supervision<br>b. Sufficient moral attention to recognize morally sensitive situation<br>c. Sufficient moral knowledge and domain knowledge to make moral decision |
| *AI requirements* | d. Machine Learning |
| *Advantages* | • Human agent is accountable for recognizing moral decision<br>• Human agent is accountable for making moral decision<br>• Artificial agent does not require moral competencies |
| *Disadvantages* | • Cognitive under- or overload of the human agent may result in missing moral choice situations or optional solutions<br>• Human agent may be turned into moral scapegoat |

*Pattern 2: Coactive Moral Decision-Maker*

In this pattern, the machine agent has more moral responsibilities than in Pattern 1. The machine agent makes the predictions regarding patients' diagnoses, and performs moral supervision on itself: it measures whether its own predictions are equally accurate for each subgroup (e.g. if diabetes diagnoses for people from Surinamese background are as accurate as for patients from other ethnic groups). The human AI developer is on stand-by. If the machine agent measures a bias in its own predictions, it initiates a handover to the human agent. In this handover, the machine agent explains to the human why a moral decision is necessary (e.g. because the model is racially biased towards people with a Surinamese background). Subsequently, the human and the computer agent make a joint decision in changing the model.

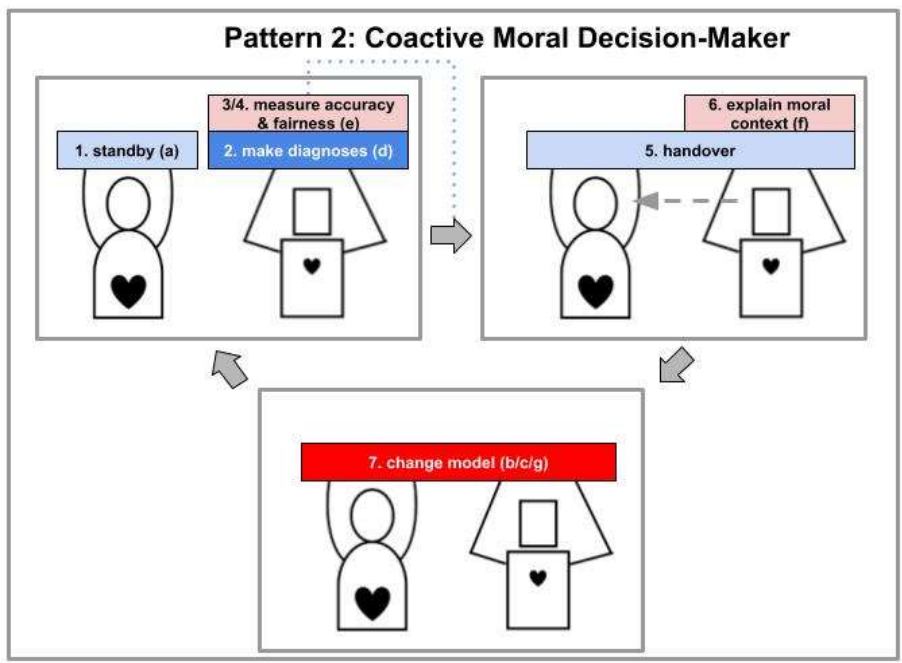

1. Human AI developer is on standby
2. Machine agent uses machine learning models to make predictions about diagnosis/treatment
3. Machine agent performs task supervision: does the model work? Are the predictions accurate?
4. Machine agent performs moral supervision: are the predictions accurate for everybody? Is there social discrimination based on subgroups?
5. If the balance between overall accuracy and fairness of the model is outside preset boundaries, the machine agent initiates a handover.
6. Machine agent explains the moral context: why is a moral decision necessary?
7. The AI developer and machine agent decide whether and how to change the model together.

| Name | 2. Coactive moral decision-maker |
|---|---|
| *Human requirements* | a. Sufficient trust in machine to recognize morally sensitive situations<br>b. Sufficient understanding of moral implications<br>c. Sufficient moral knowledge and domain knowledge to make moral decision |
| *AI requirements* | d. Machine Learning<br>e. Ability to recognize morally sensitive situations<br>f. Ability to sufficiently explain the moral context<br>g. Moral decision-support |
| *Advantages* | • Human agent is on stand-by, allowing them to do different (possibly unrelated) tasks<br>• Human agent is accountable for moral consequences, but can receive decision support by the machine agent |
| *Disadvantages* | • Morally sensitive situations that are not recognized by the machine agent are overlooked<br>• Human agent may be biased by machine's explanations and suggestions |

*Pattern 3: Autonomous Decision-Maker*

In this pattern, the machine is fully autonomous and is responsible for the entire process of making the diagnoses, assessing the fairness an accuracy, and keeping the balance by changing the model when necessary. In order to make this possible, this must be preceded by a value elicitation phase, in which the AI developer accurately programs this process. A recurrent examination (which can happen based on time passed or number of decisions made) repeats the value elicitation process and monitors possible ethical drift (Van der Waa et al., 2020). This makes sure that the machine works in accordance with the set of goals and values specified by the human.

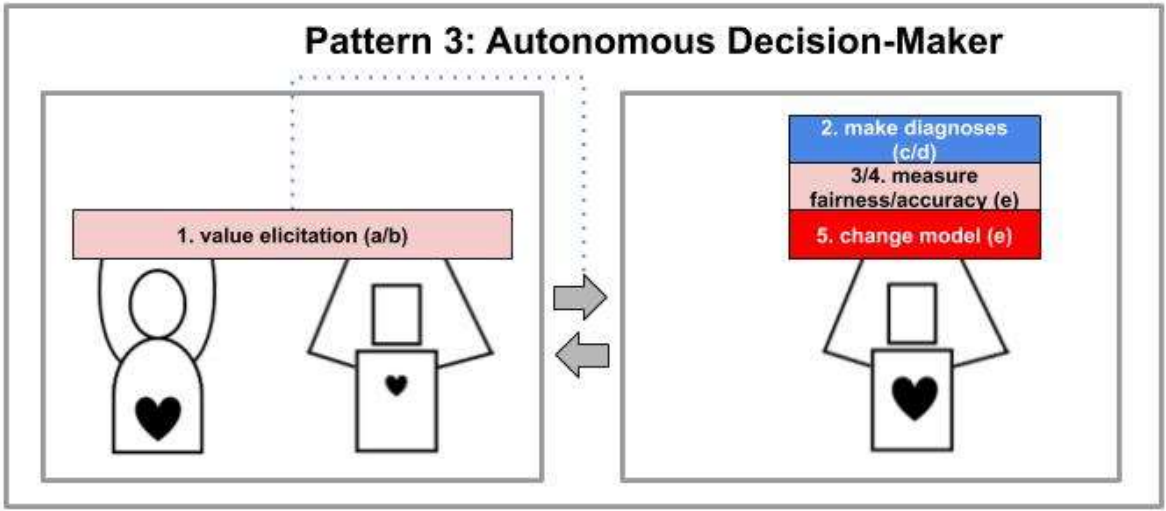

1. Human AI developer and machine agent perform value elicitation
2. Machine agent uses machine learning models to make predictions about diagnosis/treatment
3. Machine agent performs task supervision: does the model work? Are the predictions accurate?
4. Machine agent performs moral supervision: are the predictions accurate for everybody? Is there social discrimination based on subgroups?
5. If the balance between overall accuracy and fairness of the model is outside preset boundaries, the machine agent decides whether and how to change the model, in accordance with value elicitation

| Name | 3. Autonomous Decision-Maker |
|---|---|
| *Human requirements* | a. Sufficient trust in machine to recognize morally sensitive situations and handle them properly<br>b. Ability to elicitate values in machine agent |
| *AI requirements* | c. Machine Learning<br>d. Knowledge Reasoning<br>e. Ability to fully navigate moral dimension |
| *Advantages* | • Human agent is only needed for regular check-up, having time for other tasks |
| *Disadvantages* | • Value elicitation is very difficult (or even impossible) to achieve<br>• Machine agent has a lot of moral responsibility, possibly resulting in mistakes in the moral domain |

## 4.2.2 Design Challenge 1.1: 'Change the model' in Bias Mitigation

While addressing the bias mitigation design challenge described above, it became clear that other design challenges could be nested within these patterns. For example, in Pattern 2, the human and machine agent have a joint responsibility to change the model, which presents a design challenge of its own. There may be several configurations of the team-members' tasks and responsibilities to solve this sub-challenge, which may be addressed by so-called *sub-patterns*. The sections below present four such sub-patterns, describing different arrangements of the AI developer and the machine agent changing the learning model for DT2 diagnoses.

*Pattern 2.1: Memorizing Machine*

In this pattern, the machine agent has a memory support role. The human agent decides whether a model change is desirable (e.g. because the model has a strong racial bias). After this, the machine agent aids in the weighing of methods by showing previously taken measures in similar situations (e.g. over-sampling patients with a Surinamese background). The human agent takes this in consideration and picks a method to change the learning model. The machine then stores this choice in memory in order to provide future memory aid.

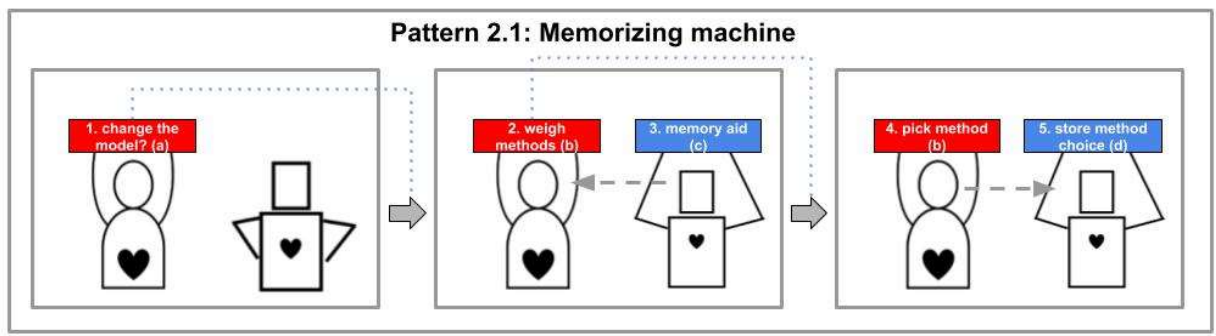

1. The human considers whether a model change is necessary. If so, the human initiates a transition.
2. The human weighs the advantages and disadvantages of the possible methods to change the learning model.
3. The machine agent provides memory aid: gives a summary of previously chosen methods in similar situations.
4. After taking this into consideration, the human picks a method to change the model.
5. The machine agent stores this choice for future memory aid.

| Name | 2.1 Memorizing machine |
|---|---|
| *Human requirements* | a. Ability to recognize and take responsibility for moral choice<br>b. Sufficient understanding of moral implications of applying different methods |
| *AI requirements* | c. Module for human memory aid<br>d. Sufficient memory for storing moral choices |
| *Advantages* | • Human is fully responsible and accountable for recognizing morally sensitive situations and making moral choice |
| *Disadvantages* | • Human cognitive under- or overload may result in mistakes<br>• Human may be incapable of predicting the consequences of all options<br>• Human agent may be biased by previously taken actions as suggested by machine's memory |

*Pattern 2.2: Learning Machine*

This pattern is similar to the previous pattern, but the machine has an additional learning mechanism. Instead of simply storing the human's previously taken choices, it creates a model of how the taken choices correlate to characteristics of the choice situation (e.g. specifics regarding the accuracy-fairness imbalance). This model is then used not for memory support, but to suggest a specific method based on previously displayed human choices.

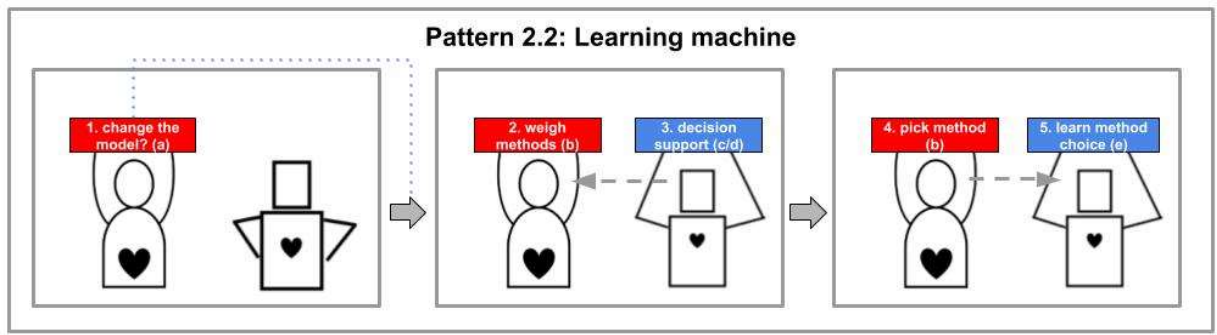

1. The human considers whether a model change is necessary. If so, the human initiates a transition.
2. The human weighs the advantages and disadvantages of the possible methods to change the learning model.
3. The machine agent provides decision support: it predicts the solution of the current problem based on a second-order model that includes previously taken choices.
4. After taking this into consideration, the human picks a method to change the model.
5. The machine agent alters its second-order model based on the human's choice.

| Name | 2.2 Learning machine |
|---|---|
| *Human requirements* | a. Ability to recognize and take responsibility for moral choice<br>b. Sufficient understanding of moral implications of applying different methods |
| *AI requirements* | c. Decision support module<br>d. Knowledge Reasoning<br>e. Machine Learning |
| *Advantages* | • Human is fully responsible and accountable for recognizing morally sensitive situations and making moral choice<br>• Decision support reduces risk of human agent's cognitive overload |
| *Disadvantages* | • Machine agent needs a large amount of data from previous choices to accurately learn the human choice behavior<br>• Machine agent's suggestions are based on human agent's previous choices, so the team may be biased towards previously taken methods (moral tunnel vision)<br>• Human may be incapable of predicting the consequences of all options |

*Pattern 2.3: Simulating Machine*

In this sub-pattern, the machine has a predictive role. As the AI developer considers the various methods possible to solve the accuracy-fairness imbalance, they can employ the machine agent to predict the effects of applying such a method. The machine simulates the effects on the fairness-accuracy balance based on previous observations and presents these to the AI developer. The human can take these simulations into account when choosing the appropriate method to change the model.

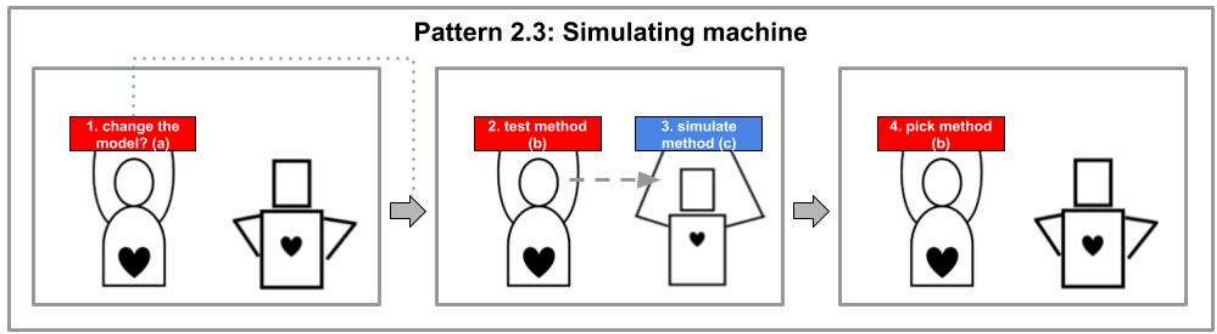

1. The human considers whether a model change is necessary. If so, the human initiates a transition.
2. The human weighs the advantages and disadvantages of the possible methods to change the learning model, and picks one to try out.
3. The machine agent simulates the effects of the method on the model's results and presents these to the human.
4. After taking this into consideration, the human picks a method to change the model.

| Name | 2.3 Simulating machine |
|---|---|
| *Human requirements* | a. Ability to recognize and take responsibility for moral choice<br>b. Sufficient understanding of moral implications of applying different methods |
| *AI requirements* | c. Capacity and computing power for accurately simulating the effect of a method on the learning model |
| *Advantages* | • Human is fully responsible and accountable for recognizing morally sensitive situations and making moral choice |
| *Disadvantages* | • Human may be incapable of predicting the consequences of all options<br>• Machine agent may not be equally capable of simulating all options<br>• Human agent may be biased towards methods that the machine agent can accurately simulate |

*Pattern 2.4: Suggesting Machine*

In this pattern, the machine has more moral responsibilities, while the human agent only has a reviewing role. In the first frame, not the human, but the machine decides whether a model change is desirable, based on preset conditions (e.g. if the learning model's diabetes diagnoses are over 10% less accurate for people with a Surinamese background). After this, the machine simulates all possible methods to change the model, and their effects on the accuracy-fairness trade-off. The machine agent then suggests the optimal method, which the human agent reviews. The human agent takes this into consideration, and finally picks the preferred method.

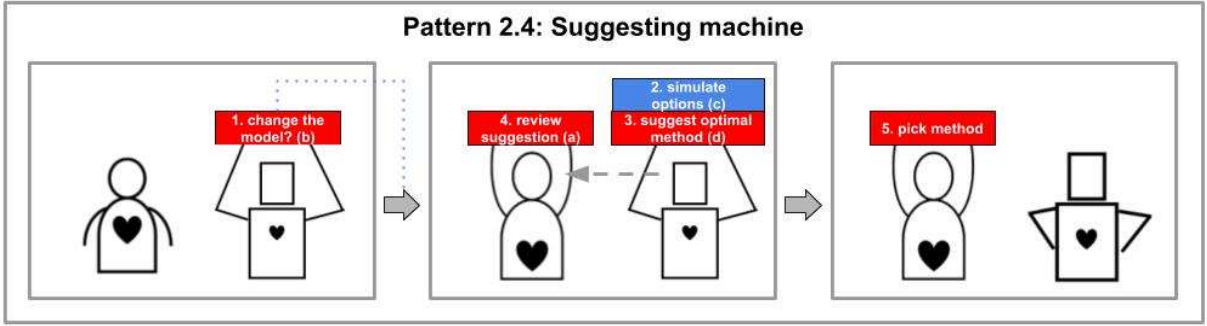

1. The machine agent considers whether a model change is necessary, based on preset rules.
2. If so, it initiates a transition. The machine agent simulates all possible methods to mitigate bias.
3. The machine agent provides decision support: it gives a suggestion of the optimal method.
4. The human reviews this suggestion.
5. The human picks the method to change the model.

| Name | 2.4 Suggesting machine |
|---|---|
| *Human requirements* | a. Sufficient trust in the machine agent's suggestions |
| *AI requirements* | b. Ability to recognize and take responsibility for moral choice<br>c. Capability to simulate the effects of all possible options<br>d. Sufficient moral understanding for picking the optimal choice |
| *Advantages* | • Low cognitive demands for human agent<br>• Clearly defined boundaries to which situations demand a moral response |
| *Disadvantages* | • High demands for machine agent's computing power<br>• Risk of machine missing morally sensitive situations if it is not included in preset boundaries<br>• Human overtrust in machine may result in little moral deliberation<br>• Machine only considers premade set of moral responses, and cannot think 'outside the box' |

### 4.2.3 Design Challenge 2: Data sharing

An important challenge in the design of medical Hybrid Intelligent systems is the sharing of patient data. In the Diabetes Type II case, there are many situations in which access to patient data can improve the accuracy and effectivity of the learning models. For example, it is desirable that the models in the system can learn if their predicted diagnosis was right or wrong. Additionally, the sharing of health-behaviors (e.g. exercise, diet, weight) can have an incredible potential for lifestyle advice personalization. Finally, for optimal medical treatment of the patient it can be very beneficial to notify the patient's doctor in the case of a critical medical situation (e.g. three hypos in one week).

However, the designers of the system have a moral and legal obligation to ask for consent to share the patient's data to their doctor or to use it to enhance the system's models. The patient's data is their property, and their autonomy may be compromised when data is shared without their consent. Yet, although it is key to respect the patient's autonomy regarding their own data, some patients may find it confronting, distressing, or simply too complex to do their research in how and to whom their data will be shared.

*Pattern 1: In-person Informed Consent*

This pattern describes a common way of attaining medical consent from a patient: through an in-person encounter with a healthcare professional (HCP). In this pattern, the HCP signs the patient into the system and explains the terms and conditions of giving consent. The HCP and the patient then perform the joint action of deciding whether to give consent to share the patient's data with the system. The HCP records this and the machine agent then shares the patient's data accordingly. This type of consent is usually all-or-nothing: there is a standard format for what the patient consents to, as there is little time to go over a large set of differentiated rules (e.g. only sharing certain data with certain individuals). The joint moral decision is largely based on a trust relationship between the HCP and the patient.

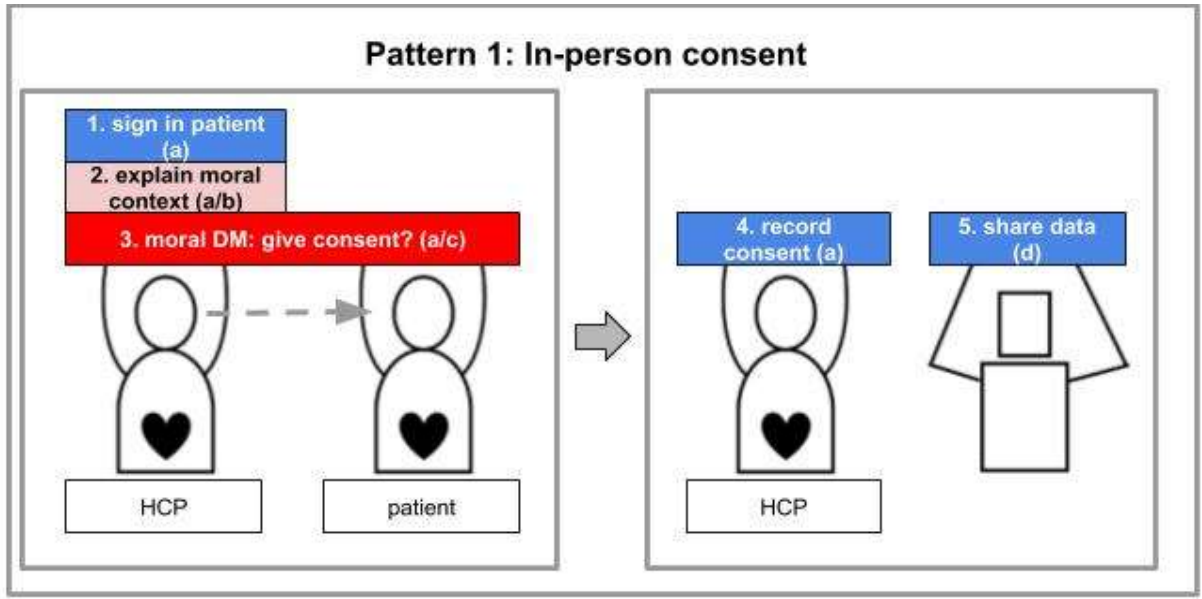

1. Healthcare professional signs in the patient in the system.
2. Healthcare professional explains the consequences of giving consent.
3. Patient and healthcare professional perform joint moral decision-making and decide whether to give consent.
4. Healthcare professional records consent in system
5. Machine agent shares data according to consent rules

| Name | 1. In-person Informed Consent |
|---|---|
| *Human requirements* | a. Health care professional's time to engage in conversation <br> b. Health care professional trained to explain consents <br> c. Trust between patient and health care professional |
| *AI requirements* | d. Simple knowledge reasoning system |
| *Advantages* | • In person conversation can ensure mutual understanding and trust <br> • Enables shared decision-making between patient and health care professional |
| *Disadvantages* | • Many possibilities for human errors <br> • Costly in terms of human time <br> • Small action space <br> • Presence of health care professional may cause interpersonal pressure to consent |

*Pattern 2: Simple Consent*

This pattern describes the configuration of simple remote consent (Whitney et al., 2003; Moore et al., 2017) between the machine agent and the patient, without the assistance of an HCP. It consists of the machine-agent signing in the patient and presenting the terms and conditions of giving all-or-nothing consent: if the patient does not consent to the predetermined set of data sharing rules, they cannot join the system or use the service. This is a very common way of obtaining consent for digital systems outside the medical realm.

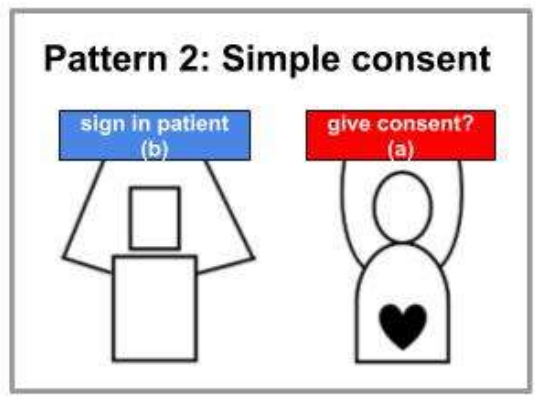

1. Machine agent signs in patient and presents the terms and conditions when giving consent.
2. Patient decides whether to give consent or not through an all-or-nothing choice.

| Name | 2. Simple Consent |
|---|---|
| *Human requirements* | a. Patient's trust in (digital) system |
| *AI requirements* | b. Simple knowledge reasoning system |
| *Advantages* | <ul><li>No missing data</li><li>Time-saving</li><li>Simple design</li></ul> |
| *Disadvantages* | <ul><li>The patient does not have the choice to (partly) join without consenting to everything (small action space)</li><li>The patient may not be aware of all the parties having access to their data</li><li>The patient may not be aware of what types of data are shared exactly</li><li>Legally weak</li></ul> |

*Pattern 3: Informed Consent*

This pattern describes a configuration of informed consent (Whitney et al., 2003), which is legally and morally stronger than simple consent. The machine agent indicates what the terms and conditions of giving all-or-nothing consent entail. Additionally, there is a mechanism through which the patient's understanding of the possible effects of giving consent is confirmed, for example through a small quiz, similar to the solution proposed by Rowbotham et al. (2013).

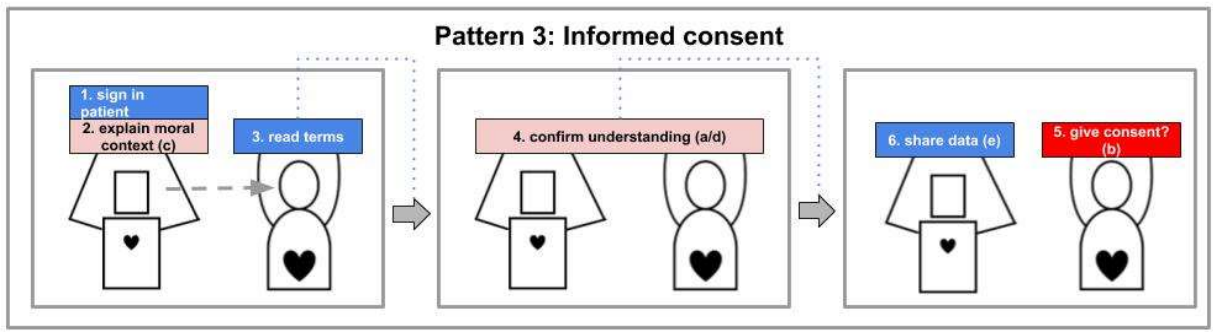

1. Machine agent signs in the patient in the system.
2. Machine agent explains the moral context (e.g. consequences) of accepting the terms and conditions.
3. Patient reads this information carefully.
4. Machine agent and patient confirm the patient's understanding of the consequences of accepting the terms and conditions.
5. Patient decides to give consent or not.
6. Machine agent shares data according to patient's consent

| Name | 3. Informed Consent |
|---|---|
| Human requirements | a. Capability and availability for understanding terms and answering questions<br>b. Patient's trust in (digital) system |
| AI requirements | c. Capability to explain consequences of giving consent<br>d. Knowledge Reasoning<br>e. Interactive module for checking whether patient understand consents |
| Advantages | • Legally stronger |
| Disadvantages | • No further support when the patient has a diminished understanding of the terms<br>• Small action space |

*Pattern 4: Differentiated Consent*

In this pattern, the patient can pick differentiated consent rules, based on the type of the shared data, and the person who the data will be visible to. In the first frame, the machine agent presents all the different consent options, explains why it is necessary that this moral decision has to be made, and what the consequences of this choice are. The patient reads the terms of these consents. In the next frame, the patient provides their consent rules, based on the type of shared data (e.g. personal, health-behaviors, medication intake, sensitive) and the person the data will be visible to (e.g. their doctor, the AI developers, no one). In the final frame, the machine agent shares the patient's data according to the rules, while the patient supervises this process. If the patient is not satisfied with one or more of the sharing rules (e.g. they realize that their weight is shared to their doctor and they do not feel comfortable with that), they can take the initiative to repeat the process and set new consent rules. This pattern is similar to the solution proposed by Rake and colleagues (2017).

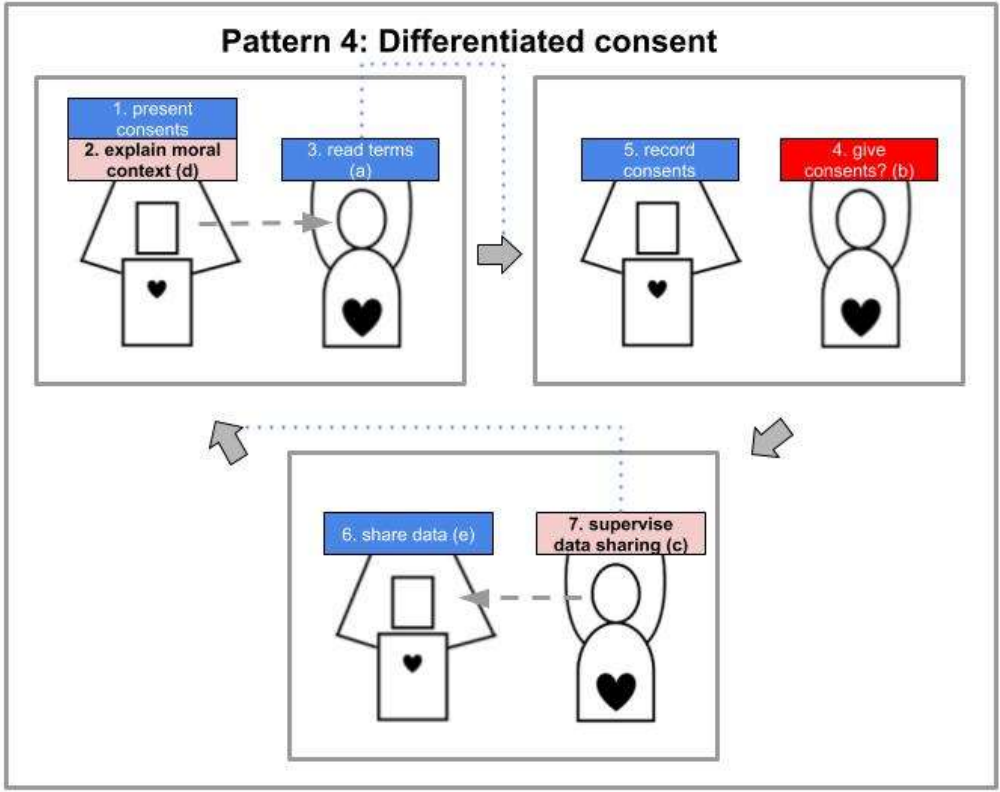

1. Machine agent presents differentiated consent rule options (for each combination of data type and receiver)
2. Machine agent explains the moral context (e.g. consequences) of certain consent rules.
3. Patient reads this information carefully.
4. Patient decides on their preferred consent rules.
5. Machine agent stores the consent rules.
6. Machine agent shares data according to the consent rules.
7. Patient supervises the machine agent in sharing their data. If desired, the patient can take the initiative to start over, and change the consent rules.

| Name | 4. Differentiated Consent |
| --- | --- |
| Human requirements | a. Patient's trust in digital system<br>b. Mental capability and availability for choosing differentiated consents<br>c. Mental availability to supervise data sharing |
| AI requirements | d. Ability to explain moral consequences of consent rules<br>e. Knowledge Reasoning |
| Advantages | • Patient has full autonomy to choose personal, differentiated consent rules |
| Disadvantages | • Results in missing data for the system and the healthcare professional<br>• Patient may be overwhelmed by choices due to high level of autonomy |

*Pattern 5: Supported Differentiated Consent*

This pattern is similar to the previous pattern but resolves the possible disadvantage of the patient being overwhelmed by choices. It starts with establishing the patient's preferred level of autonomy, for example by giving the patient three options: 'Standard consent rules', 'I want to choose but I need help', or 'Let me choose everything'. Depending on this choice, the system gives a varying number of options to choose between, accompanied with varying levels of decision-support. This way, the patients themselves can determine how complicated the consent-giving process is. Again, the patient can supervise the data-sharing process and may decide to readjust the consent rules (and, possibly, the preferred level of autonomy) if they are dissatisfied with the way their data is shared. This pattern is similar to the solution proposed by Bunnik et al. (2013).

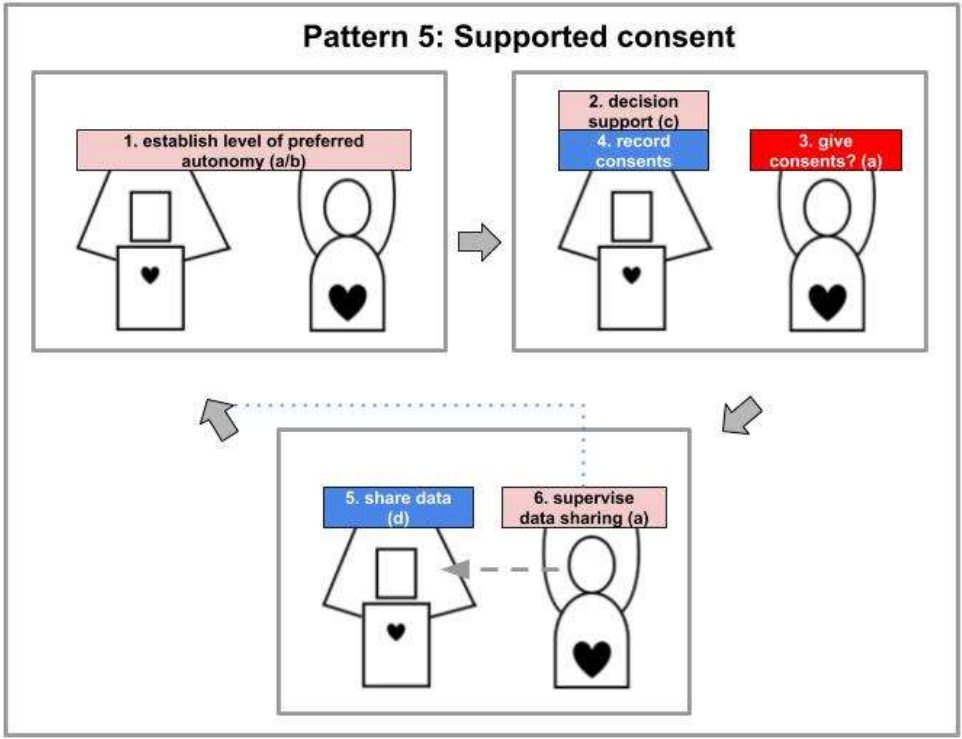

1. Machine agent and patient establish the preferred level of the patient's autonomy.
2. Machine agent provides decision support to the patient, depending on the level of preferred autonomy.
3. Patient decides on their preferred consent rules.
4. Machine agent records the consent rules.
5. Machine agent shares data according to the consent rules.
6. Patient supervises the machine agent in sharing their data. If desired, the patient can take the initiative to start over and change the consent rules.

| Name | 5. Supported Consent |
|---|---|
| Human requirements | a. Patient's trust in (digital) system |
| AI requirements | b. Interactive decision support for choosing level of autonomy<br>c. Ability to provide decision support based on preferred level of autonomy<br>d. Knowledge Reasoning |
| Advantages | • Level of autonomy is dependent on patient's capabilities and desires<br>• Large action space |
| Disadvantages | • Difficult and costly to design<br>• Results in missing data for the system as well as healthcare professional |

*Pattern 6: Self-learning Differentiated Consent*

In this pattern, the machine agent has a self-learning capability, also leading to more moral responsibilities. In the first frame, the patient chooses their consent rules, while the machine agent gives decision support proportional to how much autonomy the patient desires. When the patient provides a new consent rule, the machine agent uses this to create a model, and learn their consent behavior. In the next frame, the machine agent shares the data according to this model. When it detects a new consent type (e.g. a data type or person not yet included in the model), it switches to the third frame. In this frame, the machine agent suggests a consent rule based on its model of the patient's consent behavior. The patient then reviews this suggestion and decides whether to give consent to this rule or not. The machine agent then includes this rule in their model, and the second frame becomes active again.

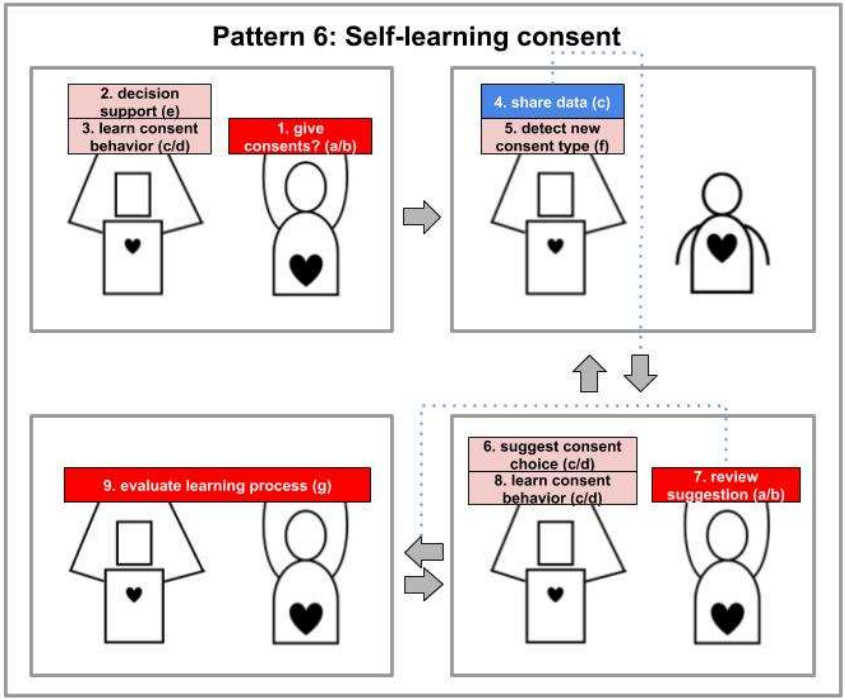

1. Human agent chooses differentiated consent rules
2. Machine agent provides decision support
3. Machine agent stores the patient's consent rules in a Knowledge Reasoning model.
4. Machine agent follows the data sharing rules.
5. Machine agent detects new types of data sharing rules that are not yet in the model (different people, data-types, or combinations)
6. Based on the model, the machine agent suggests a consent rule to the patient.
7. The human agent reviews the machine agent's suggestion and approves it or provides an alternative rule. If the human agent is displeased with the machine's suggestions, they can initiate a change of frames.
8. The machine agent adds this new rule to the model.
9. The human and machine agent evaluate the learning process, and make changes if necessary.

| Name | 6. Self-learning Consent |
|---|---|
| Human requirements | a. Capacity to understand consequences of moral consent choices<br>b. Significant trust in self-learning system |
| AI requirements | c. Knowledge Reasoning<br>d. Machine Learning<br>e. Decision support module for choosing consent rules<br>f. Ability to detect consent rules still missing from the model<br>g. Meta-learning module to evaluate and change learning process |
| Advantages | • Level of autonomy can be dependent on patient's capabilities and desires<br>• Self-learning mechanism can mitigate human cognitive underload when deciding on new consent rules that are similar to previously decided rules |
| Disadvantages | • Complex design<br>• Results in missing data for the system as well as healthcare professional |

# 5. Evaluation

## 5.0 Introduction

The previous chapter presented the Team Design Patterns that were created by combining the operational demands, human factors, and technological principles addressed in Chapter 3. The next step Socio-Cognitive Engineering method is an evaluation of the designed patterns. This is necessary to validate whether the patterns have their claimed effects and serves as input for the foundation and specification layer in the next iteration. The current chapter describes the suggested methods for this process, as there is no current standard for evaluating Team Design Patterns. The first section describes the questionnaire that was created to evaluate the patterns, as well as the qualitative and quantitative methods of analysis. The second section describes the results of using these analysis methods.

## 5.1 Methods

### 5.1.1 Data Collection

The current literature does not prescribe a standard method for evaluating Team Design Patterns. As follows from the SCE methodology, it is key to validate whether the functions in the patterns establish the desired effects as expressed in the claims. This leads to a method with four metrics: (1) understandability, (2) coherency, (3) effectiveness, and (4) generalizability. Because the patterns and pattern language are still in an early stage of development, we chose to perform a usability test with a sample of the pattern's prospective direct users: researchers and designers of hybrid intelligent systems. As the patterns are designed for the purpose of facilitating communication between different disciplines in the design process of hybrid systems, the ideal shape of the evaluation would be a focus group that simulates this process, after which participants can reflect on their experience and the patterns. However, due to time constraints of the current thesis and mobility constraints due to the COVID-19 pandemic, the research employed a questionnaire to evaluate the patterns. The motivation for using a questionnaire was that they can provide quantitative ratings regarding the metrics, but, more importantly, they can add qualitative data for a richer image of what is necessary to improve the patterns and pattern language.

*Questionnaire*

The questionnaire was created using guidelines provided by Brinkman (2009) and the Google Forms design tool. The first page of questions consisted of several five-point Likert scale questions (on a scale from *(1) very limited* to *(5) very extensive*) inquiring the participant's background knowledge in several disciplines relevant to moral decision-making in hybrid intelligent systems. The disciplines were knowledge engineering, machine learning, human cognition and behavior, interaction design, software engineering, and ethics. Additionally, participants were asked to rate their familiarity with FATE (the research program that provided the DT2 use case) and research in human-AI cooperation, Team Design Patterns, and other types of design patterns. After that, a video was presented that explained the basic elements of Team Design Patterns for moral decision-making, including the example of Jason and the robotic dog presented in Chapter 2.

The next section of the questionnaire presented the bias mitigation design challenge as described in Chapter 3, including its relevance, the actors involved, and the moral tension between beneficence and fairness it entails. It then presented two of the proposed patterns as described in Chapter 4: *The Human Moral Decision-Maker* (Pattern 1) and *The Coactive Moral Decision-Maker* (Pattern 2). The patterns consisted of a textual description combined with the pictorial and graphical representations of the patterns. The next page contained four statements for each of the two patterns, shown in Table 4. Participants were asked to what extent they agreed with these statements on a five-point Likert scale ranging from *'Completely disagree'* to *'Completely agree'*. Each of the questions was followed by the prompt '*Please explain your answer*' and a long answer text field. This section finished with an open questions regarding the two shown patterns: '*In your view, which tasks/concepts are missing in or should be added to the patterns on this page? And why?*' Afterward, participants were asked to what extent they agreed with the statement *'I understand how I'm expected to judge the patterns in this questionnaire'* on a five-point Likert scale.

| Metric | Likert scale questions for each pattern |
| --- | --- |
| Understandability | The proposed design solution of this pattern is easy to understand. |
| Coherency | The combination of pictorial and textual information in this pattern provides a coherent representation of the solution. |
| Effectiveness | The implementation of this pattern will lead to appropriate moral decision-making in human-machine diabetes care. |
| Generalizability | The solution in this pattern can be applied to other human-machine systems than diabetes care. |

**Table 4** Likert scale question for each pattern.

The next section of the questionnaire showed a video introducing the concept of 'sub-patterns', with the explanation that the joint task to *'Change the model'* in Bias Mitigation Pattern 2 can be specified further. It then presented the (sub-)design challenge of changing the model presented in Chapter 4, including its relevance, the actors involved, and the moral tension it entails (which is the same as the main pattern). It then presented two of the sub-patterns in Chapter 4: *The Memorizing Machine* (Pattern 2.1) and *The Suggesting Machine* (Pattern 2.4). Again, the patterns consisted of a textual description combined with the pictorial and graphical representations of the patterns. The succeeding page consisted of the four Likert scale statements shown in Table 4, each followed with a long text field for the participants to explain their answer. Again, the section ended with the question *'In your view, which tasks/concepts are missing in or should be added to the patterns on this page? And why?'*.

The final section of the questionnaire presented two of the data-sharing patterns described in Chapter 4: *Differentiated Consent* (Pattern 4) and *Self-learning Consent* (Pattern 6), along with the four Likert scale statements followed by open answer fields. These patterns were chosen because both are relatively complex, expectedly leading to a relatively large variance on the measures. Additionally, Pattern 4 describes an existing solution, while Pattern 6 describes a hypothetical solution that is not existent yet. Differences in responses to these patterns may provide some insight into the usability of Team Design Patterns to describe existing solutions versus their ability to plan prospective solutions to design problems.

After a pilot trial with two researchers, it was evident that filling in the questionnaire took longer than an hour. To minimize the burden on participants' schedules and to decrease the risk that they would be demotivated to start or finish the questionnaire, the final section was made optional. The rationale behind this was that, in this stage of the research, it would be better to gather the evaluations regarding fewer patterns of many researchers from different disciplines as opposed to having a few respondents evaluating more patterns.

*Participant sample*

As described above, the target users of the patterns are researchers and designers in the field of hybrid intelligence. As this evaluation is meant as a first reflection on the usability and value of the patterns and pattern language from various perspectives, it was imperative to approach researchers and designers from a broad variety of disciplines. Hence, thirty researchers and developers from various research areas at TNO were approached through a network sample. Participants were approached by email with a link to the online questionnaire, after which they had two weeks to respond. Twenty of the thirty approached researchers and designers responded, taking around 45 minutes to fill in the questionnaire. Only five respondents chose to fill in the optional part of the questionnaire. Due to this low number their quantitative results were emitted in the analysis, while their open answers were included in the qualitative analysis.

One of the approached researchers tried filling in the questionnaire but had many objections to the taken approach, resulting in limited understanding of the patterns and difficulties answering the questions. To still include the perspective of this researcher, the questionnaire was substituted with a 1.5-hour unstructured interview by phone, focusing on their vision and objections. The results were then included in the qualitative analysis.

### 5.1.2 Analysis

*Quantitative methods*

For the quantitative part of the analysis, the Likert scale questions regarding background knowledge and the ratings regarding the metrics (understandability, coherency, effectiveness, and generalizability) of each of the patterns were analyzed. Due to the nonparametric nature of Likert scale data and the relatively small sample size (N=20), the mode and median were used as an indication for the distribution of the responses. For the same reason, Spearman's rank test was performed to test for correlations between variables, while Wilcoxon's ranked sum test was used to test for statistically significant differences between ratings for the patterns.

*Qualitative methods*

The goal of the qualitative analysis is to get further insight into the metrics described above. Additionally, it aims to reveal concepts and requirements that are still missing from the patterns from the perspective of their anticipated users. Hence, the qualitative data was analyzed thematically and largely data driven. The four metrics were used as predetermined themes, in which sub-themes were inferred by categorizing the responses on an increasingly abstract level.

## 5.2 Results

### 5.2.1 Quantitative results

*Background knowledge and understanding of method*

The distributions of self-reported background knowledge are illustrated in Figure 6. Participants generally showed high self-reported experience scores for Human cognition & Behavior and Human-AI cooperation, both with a mode of 4. Previous knowledge of FATE, the research program providing the use case, was relatively low (with a mode of 1), as well as experience with Team Design Patterns (with a mode of 2).

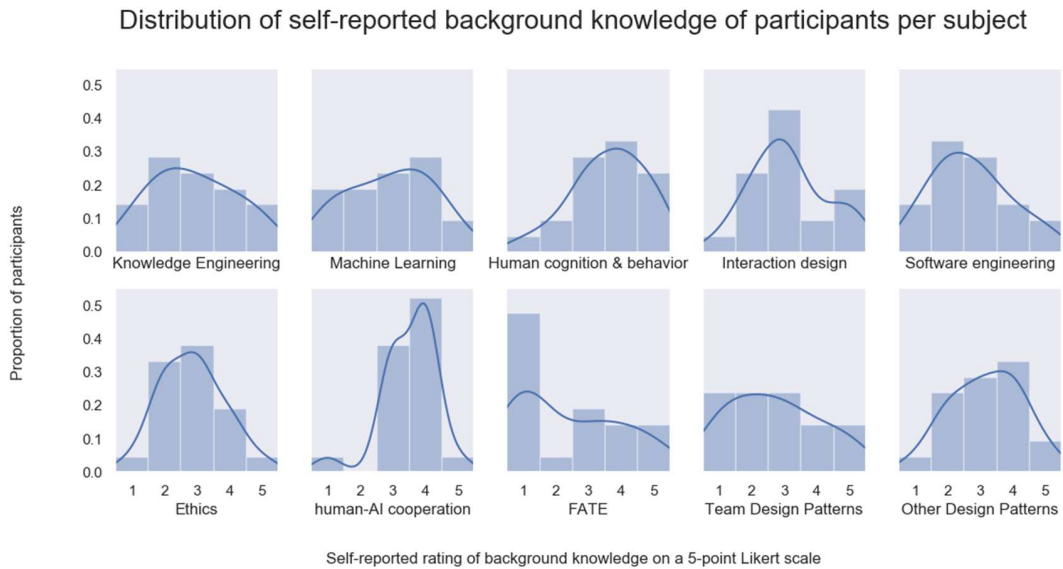

**Figure 6:** Distribution of self-reported rating of background knowledge of participants (N=20) for a range of relevant disciplines. Ratings are on a five-point Likert scale. 'FATE' refers to the research program that provides the DT2 use case of the design patterns.

Spearman tests (visualized in figure 7) showed that there was a high correlation between self-ranked expertise in software engineering and background knowledge of FATE (r=.80, p<.001). Self-ranked expertise in Machine Learning had a moderate negative correlation with reported background knowledge of Human Cognition & Behavior (r=-.63, p=.003) and Interaction Design (r=-.60, p=.005).

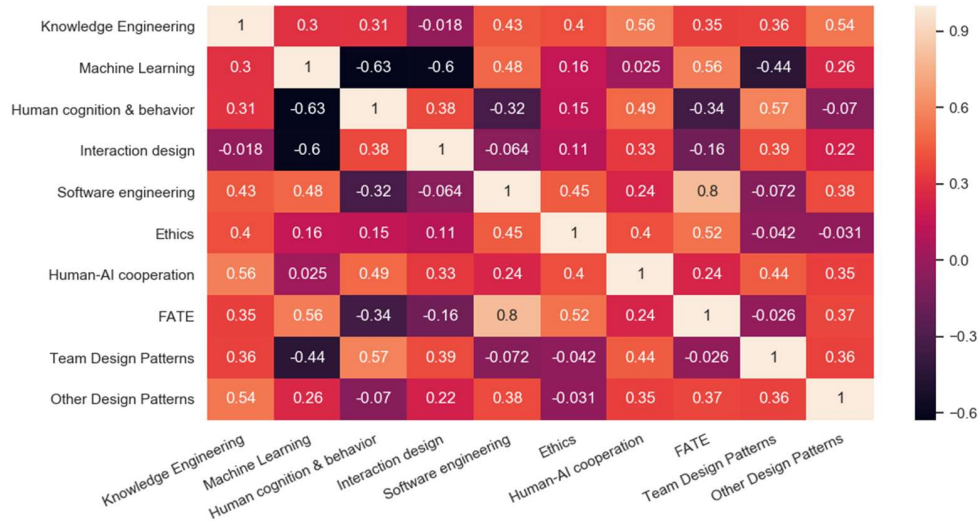

**Figure 7:** A matrix of Spearman correlation tests between all self-reported ratings regarding background knowledge of relevant disciplines.

A large majority of participants agreed with the statement *'I understand how I'm expected to judge the patterns in this questionnaire'*. No participants disagreed with this statement, while two participants responded with a neutral 3. One of these participants reported a rating of 4 on expertise in machine learning and ethics, with all ratings between 2 and 4. The other was a human cognition and behavior specialist (with a score of 5) with very low self-reported knowledge in the other research areas. Neither of these participants had prior knowledge of FATE and the Team Design Pattern approach.

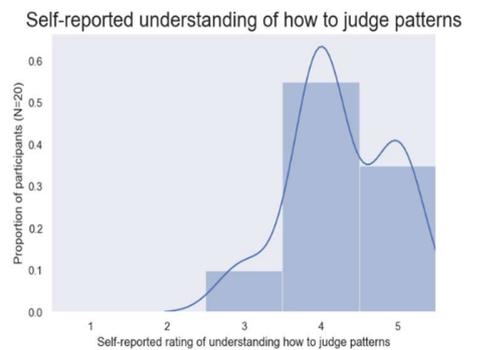

**Figure 8:** Distribution of self-reported understanding of how to judge the patterns

*Understandability*

Distributions of understandability scores of all four patterns are depicted in Figure 9. Pattern 2 was rated less understandable than Pattern 1, although not significantly (p=.055), and had a significantly lower understandability rating than sub-pattern 2.1 (p=.011) and 2.2 (p=.039). Understandability scores of Pattern 1 and Pattern 2.1 were correlated (r=.54, p=.015) and understandability scores of Pattern 2 and 2.2 were correlated (r=.48, p=.03). There were no correlations between expertise ratings and understandability scores.

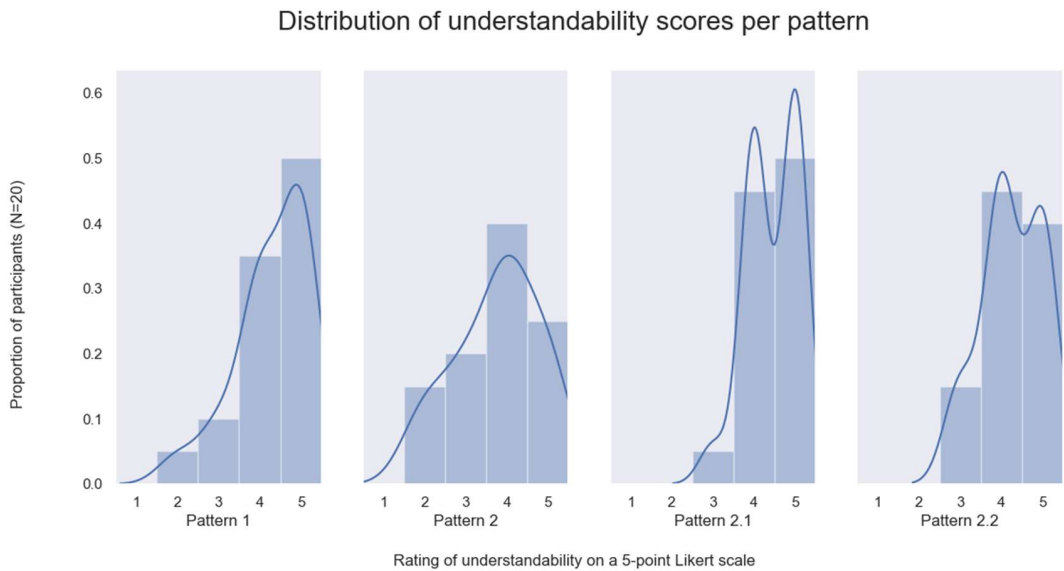

**Figure 9**: Overview of the distributions of understandability scores per pattern, rated on a five-point Likert scale.

*Coherency*

Histograms of the distributions of the patterns' coherency scores are visualized in Figure 10. Pattern 2.1 and 2.2 received significantly higher coherency scores than Pattern 1 and 2 (p<0.04). Coherency scores between all four patterns were strongly correlated (.55<r<.75, p<.02).

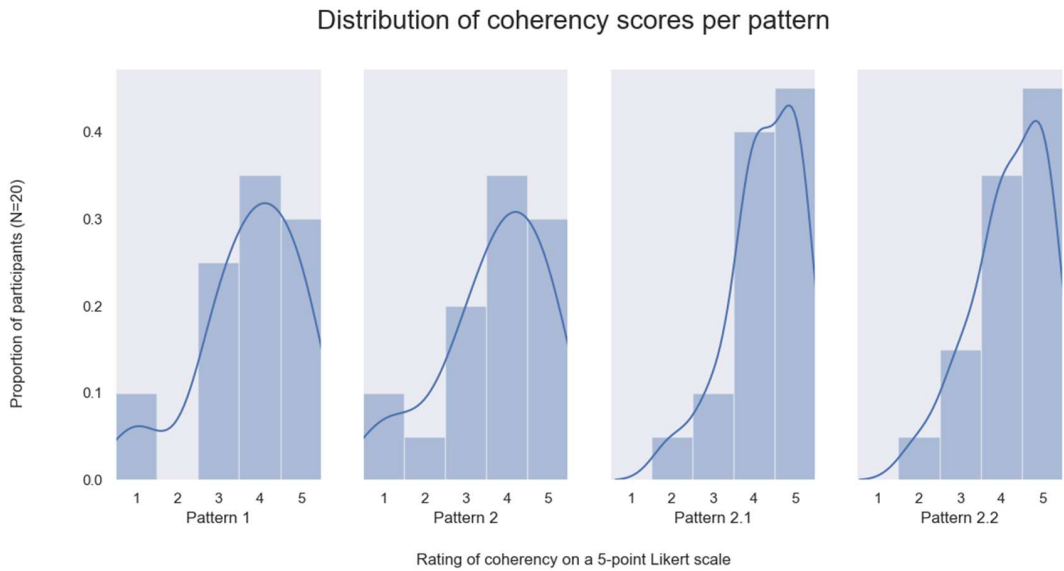

**Figure 10**: Overview of the distributions of understandability scores per pattern, rated on a five-point Likert scale.

*Effectiveness*

No significant differences were found between the responses to the statement *'The implementation of this pattern will lead to appropriate moral decision-making in human-machine diabetes care'* between any of the different patterns. Distibutions of the effectiveness scores per pattern are illustrated in Figure 11. Effectiveness scores for Pattern 1, Pattern 2.1, and Pattern 2.2 were correlated ($.45 < r < .53$, $p < .05$). Self-reported background knowledge in Knowledge Engineering showed a negative correlation with effectiveness scores of Pattern 1 ($r = -0.46$, $p = .042$) and Pattern 2.1 ($r = -0.52$ $p = .018$). Self-reported expertise in Human cognition & behavior was negatively correlated with the effectiveness scores of Pattern 2.1 ($r = -0.59$, $p = .006$). Self-reported back-ground knowledge in ethics had a negative correlation with effectiveness scores of Pattern 1 ($r = -0.49$, $p = .030$), but not with other patterns.

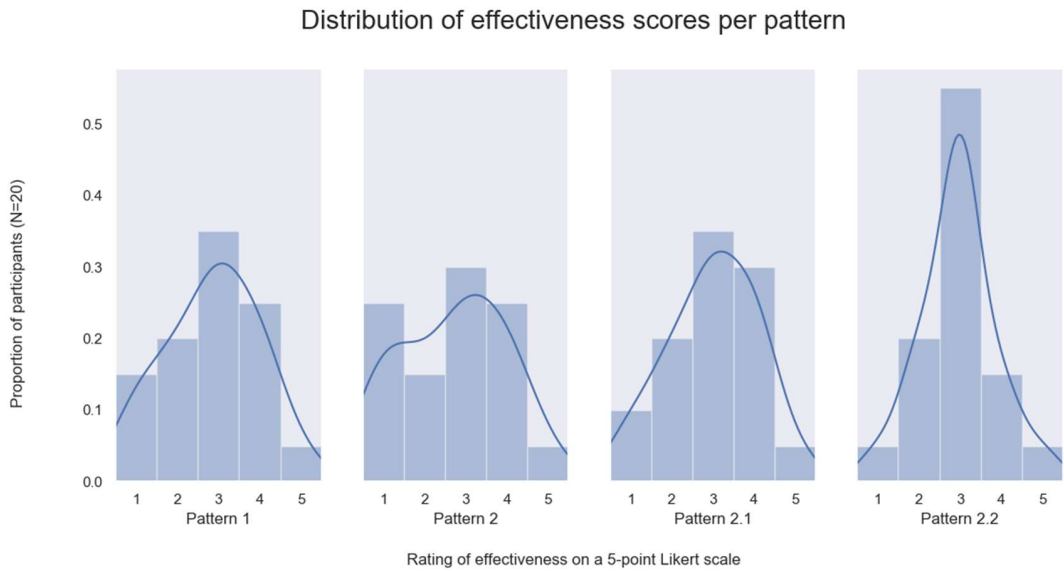

**Figure 11**: Overview of the distributions of effectiveness scores per pattern, rated on a five-point Likert scale.

*Generalizability*

The distributions of generalizability scores per pattern are depicted in Figure 12. Generalizability scores were very high, with a mode of five ('I completely agree') for all patterns with the exception of Pattern 2.2 (with a mode of 4 and 5). No significant differences were found between generalizability scores of any of the patterns. They all showed a medium to strong correlation to one another (.47<r<.70, p<0.04).

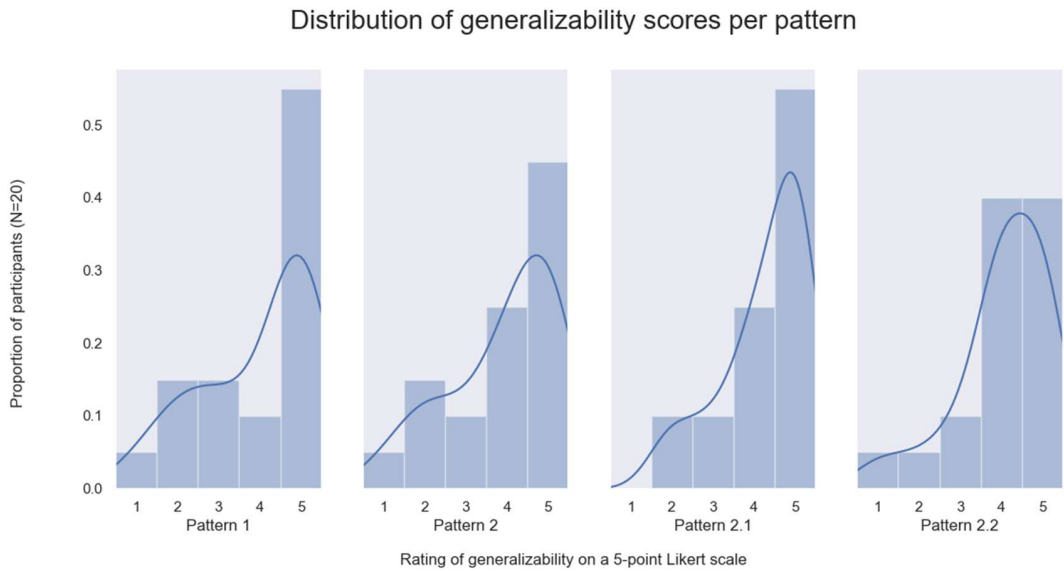

**Figure 12**: Overview of the distributions of generalizability scores per pattern, rated on a five-point Likert scale.

Finally, correlation tests were performed between the measures for each pattern. Figure 13 shows an overview of Spearman's rank correlations between pattern ratings per measure.

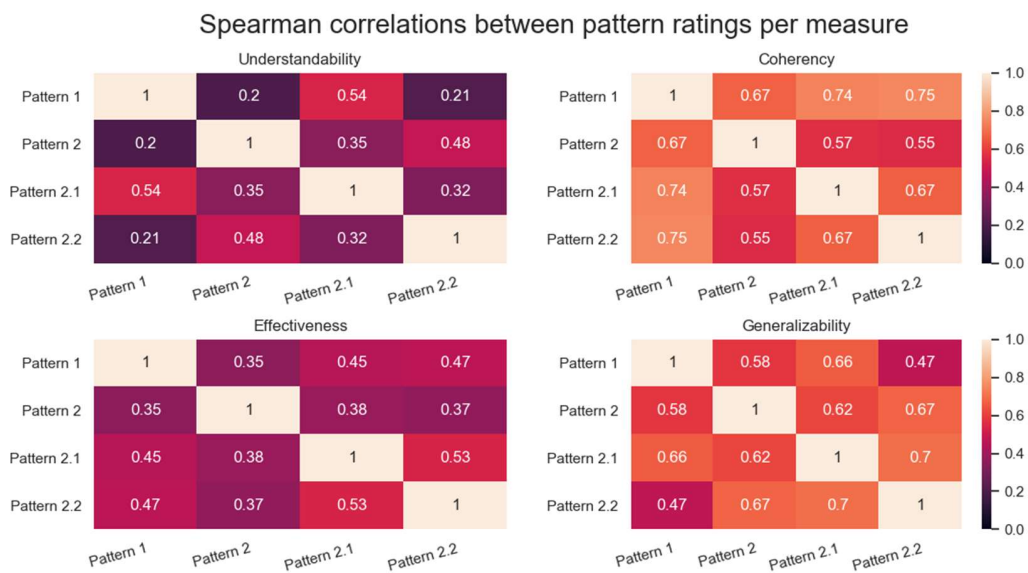

**Figure 13:** An overview of Spearman's rank correlation coefficient heatmaps of the pattern ratings per measure (understandability, coherency, effectiveness, and generalizability).

### 5.2.2 Qualitative results

*Understandability*

In the open responses of the understandability measure, three main categories of critiques were inferred. Firstly, several participants noted that they required more contextual information to fully understand the moral problem and solution presented in the patterns. For example, both Bias mitigation Pattern 1 and Pattern 2 received the remark that it was unclear how the accuracy of the system's predictions was established. One respondent mentioned that *'As I am not an expert on AI it is still a bit of a magic black box for me. It might be nice to complement the patterns with some concrete examples or explanation of some parts, e.g. what is 'the model', or what 'methods' are we talking about?'*

Secondly, several participants asked for the clarification of concepts mentioned in the pattern. One participant found the tasks in the patterns so underexposed that these patterns, or any other general or abstract type of design patterns, are too vague to solve any problem. Other respondents pointed out that concepts like 'standby' or 'together' needed to be specified in order for the patterns to be understandable. In a similar vein, a respondent remarked that some of the requirements are rather ambiguous. For example, a machine-based agent having 'sufficient moral understanding' of the consequences of an action is not only hard to imagine with the current technological principles, but also begs the difficult question when such capacity is 'sufficient'.

Thirdly, there were criticisms regarding the pictorial language. Inconsistencies in the direct-indirect task distinction and the different types of arrows were sources of some confusion, especially with regards to which agent is expected to initiate a frame switch. Additionally, several participants noted having difficulties in understanding time-related aspects of the patterns. A number of respondents found the order of the tasks confusing, as they were not always depicted clockwise or from left to right. Furthermore, one participant mentioned that sometimes sequential actions were jointly depicted in a single frame, whereas in other situations they were visualized in a new frame. Others thought it was unclear whether some tasks were continuous or recurrent.

*Coherency*

Respondents were generally very positive regarding the coherency of the design patterns. One respondent noted that *'it seems that the textual information fills in some of the subtle questions that are not in the picture'*. A large part of the participants remarked that the introductory text was vital to understanding the pattern and that the various components of the patterns were complimentary. Only one clear criticism was voiced, stating that the advantages and disadvantages in the tables were disconnected from the rest of the patterns.

*Effectiveness*

There was much variance in the responses to the effectiveness measures. Overall, multiple participants mentioned they found it difficult to judge whether the patterns would have a positive effect on the moral decision-making of the proposed system, because it would be dependent on the specific implementation. Several experts in the interaction design domain remarked that information regarding the specific interactions between the human and the system are crucial to establish the effectiveness of each of the patterns. For example, one participant stated that *'we know very little about how data and models are depicted/shown/made known to the AI developer'*, and another mentioned *'I miss how the information will be communicated'*. Respondents made it clear that the underrepresentation of this concept influenced the understandability of the patterns as well as their ability to assess the effects of their implementation.

The remainder of the remarks regarding understandability could be divided into two categories: those regarding the patterns with few machine responsibilities (Bias mitigation Pattern 1 and Pattern 2.1) and those concerning patterns with larger responsibilities for the machine agent (Bias mitigation Pattern 2 and Pattern 2.2). Most participants that were positive about the former patterns motivated their response with the fact that the human has the final say in moral issues. Conversely, participants who were more negative about these patterns remarked that the moral effectiveness depends too much on the individual human agent. An often-voiced suggestion that followed was that more humans should be involved in these patterns. Multiple respondents wrote that such moral decisions *'should not be in the hands of a single AI developer'*, advising to involve the developer's team or organization, medical experts, legislators, and insurance companies.

The responses in which more moral responsibilities are allocated to the machine (Bias mitigation Pattern 2 and 2.2) largely followed two main arguments. Firstly, there were respondents who stated that giving machines this much moral responsibility is dangerous, as it is bound to make mistakes. These participants often suggested actions or mechanisms through which the human could *'take back control'*, such as ignoring the machine's suggestions and doing nothing or picking a method that was not considered by the machine agent. A second cluster of participants viewed these patterns in a more positive light. They remarked that using these patterns would demand the *'explication of what is accurate, what is fair, etc., thus leading to more informed decision-making.'*. Thus, these respondents did not interpret the patterns as a description of a final and fully functioning system, but rather as a method to facilitate the design of a specific system.

*Generalizability*

Participants were generally very positive about the generalizability of the patterns. Those who gave high ratings to this measure noted that the patterns have 'no domain specific features' and several gave examples of other domains in which the patterns may be applicable. Some noted, however, that this *'does not mean the solution is any more correct'*, referring to their critical responses to the effectivity measure. A handful of respondents rated the generalizability score low, with the motivation that the solution described was not possible or not desirable. Two participants noted that the ability to apply the patterns to other application domains strongly depends on the moral tension in that domain. One participant wrote, for example, that *'the moral trade-off (e.g. between accuracy and fairness) may not be the same for each use case, for example because there is better input data.'*

*Missing concepts*

Finally, participants offered a wide range of suggestions in their responses to the questions which concepts were still missing from the patterns. The researcher with too many objections and difficulties to participate in the questionnaire emphasized that the legal perspective was severely lacking in the patterns. This expert in law and ethics explained that the legal realm is a structure in which society's ethical norms are democratically represented. Thus, in their view, a focus on moral decision-making in autonomous systems disregards the requirements set by the legal framework of how such systems should work. For example, the researcher mentioned that the system for diabetes care should start with the requirements and regulations as stated by the European Union's General Data Protection Regulation, after which the ambiguous moral details may be designed by an interdisciplinary team of developers. Another participant shared a milder version of this view, stating that the study of moral decision-making in hybrid systems is important, but that jurists and legislators should be included in the overview to make a connection to the legal framework.

Other respondents recommended the inclusion of more evaluation loops in the patterns. One participant suggested *'explicit AI learning from the human intervention in order to avoid it next time'*, allowing for improvement of solutions provided by the human-AI team. Additionally, several participants mentioned external evaluation loops, in which the process and results of the human-AI team are reviewed by the developer's team, organization, or an auditing committee.

Additionally, a large number of respondents mentioned that the interface between the human and machine agent is underspecified. For example, one participant stated that *'we know very little about how data and models are depicted/shown/made known to the AI developer'*, and another mentioned *'I miss how the information will be communicated'*. Participants made it clear that the underrepresentation of this concept influenced the understandability of the patterns as well as their ability to assess the effects of their implementation.

Finally, some respondents cherrypicked concepts throughout different patterns and urged to unify them in a single pattern. For example, for Data sharing Pattern 6, one participant suggested to add a mechanism through which '*the patient can still supervise/monitor the data sharing on its own initiative and make changes if desired besides only for new situations.*' This mechanism is present in Data sharing Pattern 4, to which the participant responded positively.

# 6. Discussion
## 6.0 Introduction

The previous chapter presented the quantitative and qualitative outcomes of the evaluation of the Team Design Patterns developed in this research. The current chapter examines these results. The first section discusses the combination of quantitative and qualitative results and provides a number of suggestions to improve the presented patterns and their pattern language. Section 6.2 reflects on the methodology used in this thesis and gives an overview of the limitations to the research. Section 6.3 provides concluding remarks with regards to the research questions of this thesis and the final section discusses recommendations and directions for future investigations in this field of research.

## 6.1 Discussion of results

*Background knowledge and understanding of method*

Distributions of the participants' self-reported background knowledge in the prespecified disciplines showed that (self-reported) experts from all specified research disciplines were present in the participant sample. This suggests that the sample was a good representation of the user target group: researchers and designers of hybrid intelligent systems with varying knowledge in the related disciplines. Results showed that the majority of participants had very little prior familiarity with the FATE system, indicating that no prior knowledge of the use cases is necessary to understand and assess TDPs. Only two researchers gave a neutral response to the question whether they fully understood how to judge the patterns. One was a human cognition and behavior specialist with very little knowledge of other HI fields, whereas the other had self-reported knowledge between 2 and 4 on all relevant disciplines. Hence, no clear conclusions can be drawn regarding the effect of participants' area or breadth of expertise on their ability to partake in the survey. However, a vast majority of participants indicated that they had a good understanding of what was expected of them, at least partially attesting that the questionnaire was a valid method.

*Understandability*

On the measure understandability, Pattern 2 was rated significantly lower than the other bias mitigation patterns presented. This may be explained by the fact that it was the only cyclic pattern of the patterns presented, although this is not reflected in the open answers. Instead, the qualitative data indicates that several concepts in this pattern were ambiguous, including the 'handover' from machine to human agent and the 'joint' activity of changing the model. Additionally, the interaction between the human and machine agent is more complex than in the other patterns, which makes it rather reasonable that it is slightly less understandable. Overall, the majority of respondents was positive about the understandability of all patterns.

The fact that there were no correlations between self-reported background knowledge and understandability indicates that the patterns were understandable for researchers and designers regardless of their area of expertise, which is a key requirement for their purpose of facilitating communication between disciplines. However, the qualitative data indicated that more specific examples would benefit the understanding of some of the participants with little background in AI. This resonated with other respondents, who indicated that key concepts could be specified more to optimize understandability. However, the inclusion of more information regarding more specific and concrete examples would make the patterns denser with text, which may result in less intuitive understanding.

Remarks regarding the pictorial language were mainly focused on inconsistencies. The "asymmetry arrow" was not used uniformly and may have created more confusion than understanding. Therefore, this research suggests using this arrow parsimoniously. It also recommends using clearer descriptions of the initiation arrow and the distinction between direct and indirect work, as ambiguity in these concepts may in fact hinder intuitive understanding of the patterns.

*Coherency*

Generally, participants were rather positive regarding the coherency of the different elements of the pattern. Coherency scores for the sub-patterns (Pattern 2.1 and 2.2) were significantly higher than for the more abstract bias mitigation patterns (Pattern 1 and 2). This result may suggest that lower-level patterns are to some extent easier to comprehend than more abstract patterns. However, this hypothesis is not supported by the understandability scores and should thus be taken lightly.

Coherency scores of all patterns were correlated, which may indicate that this metric is more dependent on the pattern language and less on the individual patterns, as compared to the other metrics. This is conceivable, as the texts, images and tables of all patterns consisted of the same structures and cross-references (e.g. the numbers and letters in the pictures referring to elements in the table). The remark that the advantages and disadvantages listed in the table are disconnected from the rest of the pattern is well-noted, as they are mostly hypothetical and not necessarily exhaustive. Perhaps the introductory texts of the patterns could put more emphasis on the connection between the patterns and their (dis)advantages to increase coherency. However, this thesis advises to keep including the anticipated advantages and disadvantages to encourage users to think about the effects of the patterns.

After a closer inspection of the three respondents that gave negative ratings to this measure on any of the patterns, it seemed that these participants interpreted the statement '*The combination of pictorial and textual information in this pattern provides a coherent representation of the solution*' differently than anticipated. These participants were critical of the word 'solution', as they thought the effects of the design solution would be negative and hence would not solve the moral problem. Future work should ensure emphasize that this measure concerns coherency and not whether the solution is a very good one.

*Effectiveness*

Effectiveness scores of all patterns were low compared to the other metrics, with a mode of 3 for all patterns. Part of this may be explained by noise in the data caused by the slightly ambiguous phrasing of the question: it may not have been clear to the participants what 'appropriate moral decision-making' meant. One may argue that this phrasing left room for interpretation of each participant, drawing on their experience and expertise to make an intuitive assessment whether a proposed solution would improve the system. However, to increase interpretability of the data, future research may deconstruct the effectiveness measure into simpler and clearer questions (e.g. 'This pattern facilitates mutual understanding between researchers from various relevant backgrounds', 'This pattern would lead to a more fair system', and 'This pattern would lead to more explicit consideration of moral decision-making in the design process').

As indicated in the qualitative analysis, many participants expressed that they found it hard to judge whether the use of the patterns would really result in better moral decision-making because this depends on the specific implementation. It is conceivable that these doubts moved participants to give more 'neutral' ratings for this measure. The criticism to make the patterns more specific resonates with an important tension that runs through the core of Team Design Patterns, between their requirement to be generalizable, abstract, intuitive and uncluttered on the one hand, and the desire to include ample detailed information to make them applicable to specific situations on the other. In a similar vein, interaction design experts indicated that the way information is communicated between human and machine agent has a strong impact on how implementation of the patterns would play out. While this is truthful, it is not the aim of Team Design Patterns to give a detailed description of these interactions, as there is a widely accepted methodology of design patterns available for this (Borchers, 2008). An important focus point of future research is the integration and interaction of such Interaction Design Patterns (IDPs) and Team Design Patterns. The combination of these two types of patterns that operate on two different abstraction levels would preserve the strong assets of both: the generalizability and focus on responsibility of TDPs, and the detailedness and readiness for application of IDPs.

A large part of the participants was positive about the patterns in which humans clearly had the final say, which resonates with the concept of meaningful human control. Conversely, some respondents noted that some of the patterns were too dependent on human individuals. Interestingly, expertise in ethics was negatively correlated with ratings of the effectiveness of Pattern 1, but not of other patterns. This may be explained by the worry that the human may perceive cognitive under- or overload, resulting in worse moral decision-making. Ethicists may be particularly prone to this worry, as they are aware of the human (in)capabilities in the moral sphere. Alternatively, this finding may be explained by the fact that this pattern was the first to be presented, when the participants may be relatively more critical regarding certain features because they are 'new'. One of these features may be the fact that the AI developer is the only human involved in any of the design solutions for bias mitigation. An often-voiced criticism was that other humans, such as the developing team, organization, or auditing committee should be involved in the process in order for the patterns to truly lead to improved moral decision-making. This is a very valid critique, but the inclusion of more actors in the patterns may be at the expense of the intuitiveness and understandability of the patterns, as it quickly leads to cluttered images. Perhaps the addition of not new actors, but new tasks could resolve this issue (e.g. *'talk to supervisor'*, *'send notice to auditing committee'*, etc.).

*Generalizability*

Participants were largely positive regarding the generalizability of the patterns. Additionally, generalizability scores were mostly stable regardless of the specific pattern. These results underwrite that generalizability and reusability is one of the main strengths of the TDP methodology, as indicated by Van Diggelen et al. (2018). Similarly to the coherency scores, qualitative results indicated that some of the respondents interpreted this measure's statement differently than intended, noting that they disagreed with the proposed solution. Future work may take this into account by making clear that this measure does not concern the quality of the proposed method.

## 6.2 Limitations

There were several limitations to the current research. Most importantly, time constraints, as well as the onset of the COVID-19 pandemic, influenced the methods used for the gathering of data. The operational demands were analyzed in an exploratory fashion, drawing from various sources including project team members, while a questionnaire was used for evaluating the patterns. However, as was mentioned in section 5.1, focus groups would have been more appropriate for two reasons. Firstly, a focus group for the operational demands would have likely resulted in more specific demands for the patterns towards their prospective users. Secondly, a focus group in which various designers and researchers would jointly use and reflect on the patterns would result in more thorough usability data.

Secondly, due to time constraints it was necessary to focus on two out of three design challenges as identified in Chapter 3.1. The choice to focus on bias mitigation and data sharing was motivated by the fact that these challenges are fairly different, which allowed the evaluation of the Team Design Pattern methodology in different applications. However, the inclusion of the 'lifestyle intervention' challenge would likely have added valuable insights into the workings of the patterns. Specifically, this design challenge has a moral tension between the patient's autonomy and beneficence of the system, likely resulting in a consent-giving mechanism in its solutions, similar to the data sharing design challenge. Creating and comparing Team Design Patterns for both these challenges may have resulted in increased understanding of the generalizability of the patterns.

Additionally, the tool to create the survey provided very limited means to change the layout. This influenced the assessment method, as it was difficult for participants to compare patterns. Ideally, design patterns for one design problem would have been positioned not below but next to each other. Future research in this field may consider using a more advanced tool than Google Forms, or examine the development of a tool specifically geared towards the testing of Team Design Patterns.

Another limitation to this study is that all participants were researchers in the same research institute, possibly leading to a selection bias. Although the data showed that many participants had little previous experience with FATE or Team Design Patterns, there are no guarantees that understanding of the patterns may rely on background knowledge in another domain that was not included in the background inquiry. Researchers in the research institute may even have (subconscious) culture-dependent know-how that provides a foundation for the understanding or inclination towards the design patterns. Furthermore, although several significant results were found, a larger participant sample would likely have resulted in more robust quantitative findings, especially with the use of nonparametric significance testing (Sullivan, 2013). The inclusion of more experts from each of the specified research disciplines would enable a better comparison of the quantitative measures between groups of researchers.

A final limitation to this research is that not all created patterns could be evaluated. It took the researchers 30-60 minutes to fill in the questionnaire regarding four of the patterns, while they were not compensated for their participation. Only five of the participants chose to continue with the data sharing patterns, indicating that the inclusion of even more patterns was infeasible in the current setup. Nonetheless, the small selection of patterns may have created a false dilemma: participants felt like they could only choose between the two provided options (e.g. little machine responsibility or much machine responsibility). Some of the respondents advocated combining the assessed patterns, or suggested mechanisms that were in the patterns not included in the evaluation. This phenomenon may have led to a loss of nuance. However, the goal of the current research was not to provide an exhaustive overview of which of the patterns was best, but rather to compare the features in the patterns to determine directions for forthcoming research.

## 6.3 Conclusions

These limitations considered, several conclusions can be drawn on the basis of this research. The Socio-Cognitive Engineering methodology was used to combine state-of-the-art knowledge of human factors, available technologies, and operational demands in order to create a set of patterns for the design problems of bias mitigation and the sharing of health-related patient data. A questionnaire was developed that was used to evaluate the usability of these patterns. The results provided the following insights to the research questions specified in Chapter 1.3:

*RQ 1a: How understandable are the created Team Design Patterns for their potential users?*

The results showed that participants had a good comprehension of the inner relations and functioning of the patterns, as they found that the combination of textual, graphical, and tabular features constituted design patterns had a high degree of coherency. Understanding of the content of the proposed design solutions were more dependent on the specific pattern, but overall results showed that the created TDPs were largely understandable to the wide range of experts involved in the design of HI systems.

*RQ 1b: How generalizable are the created Team Design Patterns?*

Expert evaluations indicated that the patterns had no domain-specific features, rendering them reusable for similar design problems in- and even outside the medical domain. Answering this research question, results showed that the created Team Design Patterns were largely generalizable. Future research may give more concrete answers to this question by testing the application of these design patterns to different situations.

*RQ 1c: To what extent does the use of the created Team Design Patterns lead to more appropriate moral decision-making?*

Results of the evaluation were mostly neutral regarding the effectivity of the provided design patterns. Qualitative data provided several substantive suggestions to make sure that the patterns would lead to better moral decision-making, including interaction design and the involvement of additional actors. However, qualitative results also showed that the evaluation setup was not fully suitable to assess whether the use of these patterns leads to more appropriate moral decision-making than without heir use, as participants required more information regarding the specific configurations. Several respondents, however, noted their belief that the use of these design patterns would lead to more explicit discussions about moral decision-making in the design process, ultimately leading to better-functioning systems.

> *RQ 1: How can Team Design Patterns describe moral decision-making for bias mitigation and data sharing in a medical hybrid intelligent system, so that they are usable by researchers from the various disciplines involved in the design of such systems?*

All in all, the answers to the sub-questions provided above provide a mixed but promising answer to this main research question: The Team Design Patterns presented in this thesis exemplify descriptions of design solutions for bias mitigation and data sharing that are generalizable and understandable for researchers from the various disciplines involved in the design of medical HI systems. Future research should be directed towards uncovering whether and how these patterns can result in improved moral decision-making in their application. Additionally, several concepts are still missing from these patterns that may be incorporated to attain more complete descriptions.

> *RQ 2: Which methodological tools are suitable for the creation of such Team Design Patterns?*

Considering all the answers above, the SCE methodology has proven to be suitable for the creation of TDPs. It is specifically geared towards the integration of human factors knowledge and technological principles for specific operational demands, providing a useful structure for creating patterns that have a Hybrid Intelligent nature. The incorporation of Value-Sensitive Design elements led to explicit consideration of stakeholder values from the very start of the design process, naturally resulting in the inclusion of the moral dimension in the design patterns. Participants generally had a good understanding of how they were expected to partake in the questionnaire that was developed for the evaluation of the patterns. Results suggest that the survey was a useful tool to assess the understandability, generalizability, and, to a smaller extent, the effectivity of the created patterns. SCE's cyclic character allows for incremental improvements of both the patterns and evaluation method based on new research in the human factors and technological domains or the patterns' application to new use cases or domains.

Overall, this research has provided reusable conceptualizations for solving the bias mitigation and data sharing design problems from a Hybrid Intelligence perspective. It demonstrated that the TDP language is suitable for the unambiguous description and communication of design problems and their (hypothetical) solutions. This was possible for different design problems on varying degrees of abstraction. The research illustrated that TDPs can be created and applied in the design process of actual HI systems and showed that the created patterns were usable for researchers and designers from a broad range of relevant disciplines.

## 6.4 Recommendations

As indicated in section 6.2, future work in this field of research can make several methodological improvements for enhanced usability testing. Most importantly workshops or focus groups may aid in getting a deeper insight into the usability of the patterns by providing interactive and behavioral data. Additionally, the effectiveness measure can be deconstructed to make the assessment more understandable and prevent noise in the usability data.

The TDP language would benefit considerably by the inclusion of several concepts. Particularly, references to the interaction design would likely make the patterns more understandable and concrete. Connections to legal requirements would draw the users' attention to laws that may be important for the systems' design. The inclusion of evaluation loops and the possibility to go "outside" the pattern and communicate with external actors (development team, organization, auditing committee, legislator, patient's family, etc.) may lead to improved moral decision-making. This resonates with a body of human factors literature stating that conversation and evaluation are a fundamental aspect of moral decision-making (see section 3.2.2).

Additionally, future work may focus on the development of a tool specifically geared towards creating team design patterns. The Google Draw application was sufficient for the current research, but a specialized tool could aid quicker and more flexible constructions of TDPs. This would not only make it easier to develop and assess TDPs, but, more importantly, this would allow researchers from various disciplines to create and adapt their own TDPs in the design process of HI systems. A specialized tool for the creation of the patterns is crucial for a universal adoption of the TDP language and the building of a library of reusable design patterns.

Another line of research that would greatly contribute to the development of ethical HI systems is the study of AI requirements in the TDP language. The current research addressed high-level functions the machine agent should have to take various support roles, including 'the ability to explain moral consequences' and 'the ability to recognize morally sensitive situations.' A lot of progress is yet to be made in formalizing these functions. Which methods are necessary for this, and what metrics should be used to measure success? Detailed taxonomies in these domains as well as the patterns' descriptions in (pseudo-)code may be first steps in the formalization and application of these patterns.

The development of the Team Design Pattern language is an ambitious project that can structure today's plethora of academic and technological developments and substantially assist the design of HI systems with growing complexity. Their main pitfall, however, is that they have little guarantee of completeness as they inherently touch upon a tremendous spectrum of academic concepts that are impossible to do justice in a single project. However, this is exactly the strength of both TDPs and the SCE methodology: they allow for careful documentation and continuous improvement based on their application to new domains. The results of this thesis have provided a clear advancement in the development of Team Design Patterns for moral decision-making in medical HI systems in general and for the Bias Mitigation and Data Sharing design problems specifically. This can function as a benchmark for future research into moral decision-making in HI systems, furthering the agenda of creating configurations of an ethical synergy between humans and technology.